\documentclass[twocolumn,tighten,twocolappendix]{aastex701}
\usepackage{wrapfig}
\usepackage{graphics,graphicx}
\usepackage{epstopdf}
\usepackage[latin1,utf8]{inputenc}
\usepackage{tikz}
\usepackage{booktabs}
\usetikzlibrary{shapes,arrows}
\usepackage{amssymb, amsmath, amsthm}
\usepackage[normalem]{ulem}
\usepackage{soul} 
\usepackage{float}
\usepackage{color}
\usepackage{xcolor}
\definecolor{ultramarine}{rgb}{0.01, 0.64, 0.86}
\definecolor{deepgreen}{RGB}{0,100,0}

\begin{document}

\title{Tails of Gravity: Persistence of Star Formation in the CMZ Environment}

\correspondingauthor{Hauyu Baobab Liu, Sihan Jiao}
\email{baobabyoo@gmail.com, sihanjiao@nao.cas.cn}

\author[orcid=0009-0003-4821-5502]{Linjing Feng}
\affiliation{National Astronomical Observatories, Chinese Academy of Sciences, 20A Datun Road, Chaoyang District, Beijing 100012, China}
\affiliation{University of Chinese Academy of Sciences, Beijing 100049, China}
\email{ljfeng@nao.cas.cn}

\author[orcid=0000-0002-9151-1388]{Sihan Jiao}
\affiliation{National Astronomical Observatories, Chinese Academy of Sciences, 20A Datun Road, Chaoyang District, Beijing 100012, China}
\affiliation{Max Planck Institute for Astronomy, Konigstuhl 17, D-69117 Heidelberg, Germany}
\email[]{sihanjiao@nao.cas.cn}

\author[0000-0001-5950-1932]{Fengwei Xu}
\affiliation{Kavli Institute for Astronomy and Astrophysics, Peking University, Beijing 100871, People's Republic of China}
\affiliation{Department of Astronomy, School of Physics, Peking University, Beijing, 100871, People's Republic of China}
\affiliation{I. Physikalisches Institut, Universität zu Köln, Zülpicher Str. 77, D-50937 Köln, Germany}
\email{}

\author[0000-0003-2300-2626]{Hauyu Baobab Liu}
\affiliation{Department of Physics, National Sun Yat-Sen University, No. 70, Lien-Hai Road, Kaohsiung City 80424, Taiwan, R.O.C.}
\affiliation{Center of Astronomy and Gravitation, National Taiwan Normal University, Taipei 116, Taiwan}
\email[]{baobabyoo@gmail.com}

\author[]{Xing Lu}
\affiliation{Shanghai Astronomical Observatory, Chinese Academy of Sciences, 80 Nandan Road, Shanghai 200030, PR China}
\email[]{xinglu@shao.ac.cn}

\author[]{Neal J. Evans II}
\affiliation{Department of Astronomy, The University of Texas at Austin, 2515 Speedway, Stop C1400, Austin, 78712-1205, USA}
\email[]{nje@astro.as.utexas.edu}

\author[0000-0001-8782-1992]{Elisabeth A.C. Mills}
\affiliation{Department of Physics and Astronomy, University of Kansas, 1251 Wescoe Hall Drive, Lawrence, KS 66045, USA}
\email[]{}

\author[0000-0001-8991-9088]{Attila Kov\'{a}cs}
\affiliation{Center for Astrophysics $|$ Harvard \& Smithsonian, 60 Garden Street, Cambridge, MA, 02138, USA}
\email{}

\author[]{Qizhou Zhang}
\affiliation{Center for Astrophysics $|$ Harvard \& Smithsonian, 60 Garden Street, Cambridge, MA, 02138, USA}
\email{}

\author[0000-0001-9299-5479]{Yuxin Lin}
\affiliation{Max-Planck-Institut f\"ur Extraterrestrische Physik, Giessenbachstr. 1, D-85748 Garching bei M\"unchen, Germany}
\email{ylin@mpe.mpg.de}

\author[]{Jingwen Wu}
\affiliation{University of Chinese Academy of Sciences, Beijing 100049, China}
\affiliation{National Astronomical Observatories, Chinese Academy of Sciences, 20A Datun Road, Chaoyang District, Beijing 100012, China}
\email{jingwen@nao.cas.cn}

\author[0000-0002-9390-9672]{Chao-Wei Tsai}
\affiliation{National Astronomical Observatories, Chinese Academy of Sciences, 20A Datun Road, Chaoyang District, Beijing 100012, China}
\affiliation{Institute for Frontiers in Astronomy and Astrophysics, Beijing Normal University,  Beijing 102206, China}
\affiliation{University of Chinese Academy of Sciences, Beijing 100049, China}
\email{cwtsai@nao.cas.cn}

\author[]{Di Li}
\affiliation{New Cornerstone Science Laboratory, Department of Astronomy, Tsinghua University, Beijing 100084, China}
\affiliation{National Astronomical Observatories, Chinese Academy of Sciences, 20A Datun Road, Chaoyang District, Beijing 100012, China}
\email{dili@mail.tsinghua.edu.cn}

\author[]{Zhi-Yu Zhang}
\affiliation{School of Astronomy and Space Science, Nanjing University, Nanjing 210093, China}
\affiliation{Key Laboratory of Modern Astronomy and Astrophysics, Ministry of Education, Nanjing 210093, China}
\email{zzhang@nju.edu.cn}

\author[]{Zhiqiang Yan}
\affiliation{School of Astronomy and Space Science, Nanjing University, Nanjing 210093, China}
\affiliation{Key Laboratory of Modern Astronomy and Astrophysics, Ministry of Education, Nanjing 210093, China}
\email{yan@nju.edu.cn}

\author[]{Hao Ruan}
\affiliation{University of Chinese Academy of Sciences, Beijing 100049, China}
\affiliation{National Astronomical Observatories, Chinese Academy of Sciences, 20A Datun Road, Chaoyang District, Beijing 100012, China}
\email{ruanhao@bao.ac.cn}

\author[]{Fangyuan Deng}
\affiliation{University of Chinese Academy of Sciences, Beijing 100049, China}
\affiliation{National Astronomical Observatories, Chinese Academy of Sciences, 20A Datun Road, Chaoyang District, Beijing 100012, China}
\email{dengfangyuan21@mails.ucas.ac.cn}

\author[]{Yuanzhen Xiong}
\affiliation{University of Chinese Academy of Sciences, Beijing 100049, China}
\affiliation{National Astronomical Observatories, Chinese Academy of Sciences, 20A Datun Road, Chaoyang District, Beijing 100012, China}
\email{}

\author[]{Ruofei Zhang}
\affiliation{University of Chinese Academy of Sciences, Beijing 100049, China}
\affiliation{National Astronomical Observatories, Chinese Academy of Sciences, 20A Datun Road, Chaoyang District, Beijing 100012, China}
\email{zhangruofei24@mails.ucas.ac.cn}

\begin{abstract}
\setlength{\parindent}{0pt}
We characterize star-forming gas in six molecular clouds (Sgr~B1-off, Sgr~B2, Sgr~C, the 20 km\,s$^{-1}$ and 50 km\,s$^{-1}$ molecular clouds, and the Brick) in the Galactic central molecular zone (CMZ), and compare their star-forming activities with those in molecular clouds outside the CMZ.
Using multi-band continuum observations taken from {\it Planck}, {\it Herschel}, JCMT/SCUBA-2, and CSO/SHARC2, we derived 8\farcs5 resolution column density maps for the CMZ clouds and evaluated the column density probability distribution functions (N-PDFs).
With the archival Atacama Large Millimeter/submillimeter Array (ALMA) 1.3 mm dust continuum data, we further evaluated the mass of the most massive cores ($M_{\rm core}^{\rm ma x}$).
We find that the N-PDFs of four of the selected CMZ clouds are well described by a piecewise log-normal$+$power-law function, while the N-PDFs of the remaining two can be approximated by log-normal functions.
In the first four targets, the masses in the power-law component ($M_{\rm gas}^{\rm bound}$), $M_{\rm core}^{\rm max}$, and star formation rate (SFR) are correlated.
These correlations are very similar to those derived from low-mass clouds in the Solar neighborhood and massive star-forming regions on the Galactic disk.
These findings lead to our key hypotheses: (1) In the extreme environment of the CMZ, the power-law component in the N-PDF also represents self-gravitationally bound gas structures, and (2) evolution and star-forming activities of self-gravitationally bound gas structures may be self-regulated, insensitive to the exterior environment on $\gtrsim$5--10 pc scales.

\keywords{Star formation(1569) --- Initial mass function(796) --- Molecular clouds(1072) --- Galactic center(565)}

\end{abstract}

\section{Introduction}

\begin{figure*}[t!]
    {\renewcommand{\arraystretch}{0}
    \hspace{-0.6cm}\begin{tabular}{ p{1cm} }
        \includegraphics[width = 17.5cm]{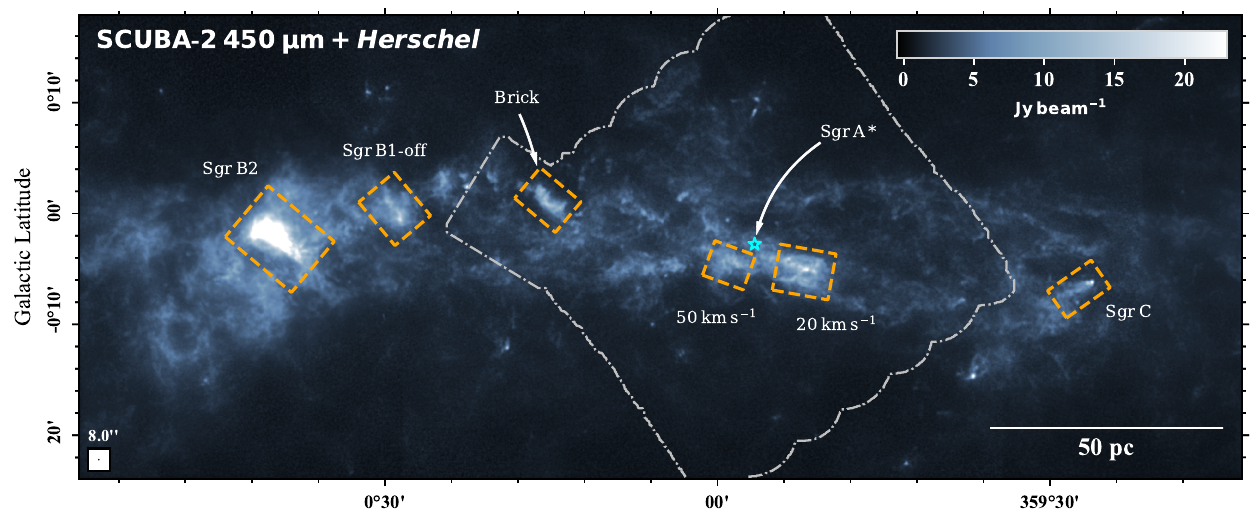}\\
        \includegraphics[width = 17.5cm]{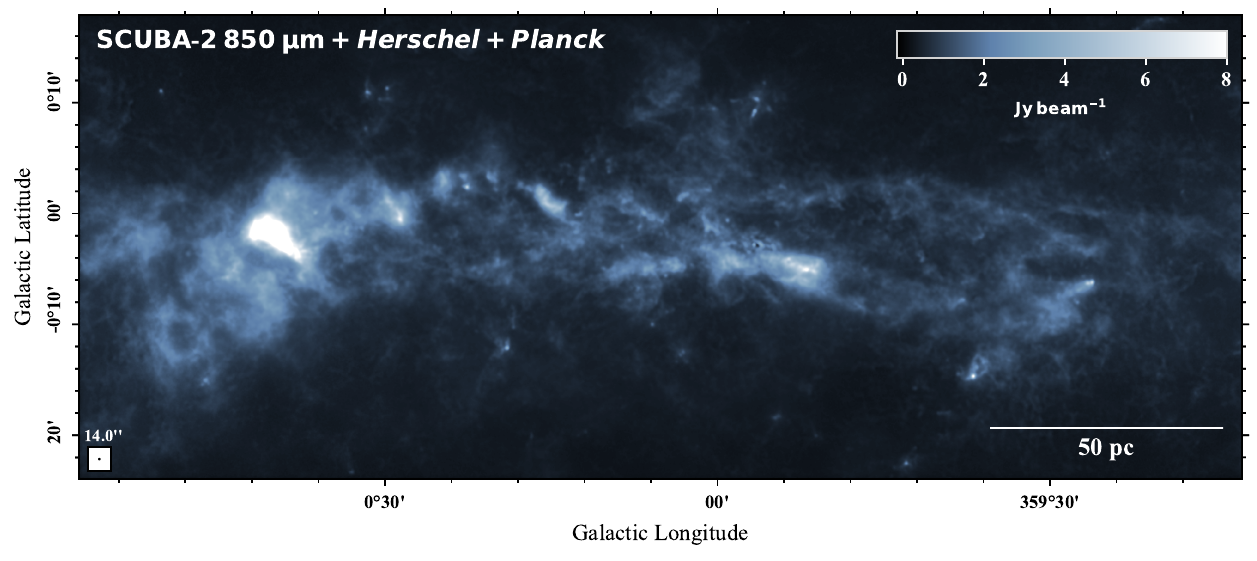}\\
    \end{tabular}
    }
    \caption{
    The 450 ${\rm \mu m}$ (top panel) and 850 ${\rm \mu m}$ (bottom panel) submillimeter maps of the Central Molecular Zone, produced by combining {\it Herschel}/{\it Planck} and JCMT-SCUBA2 observations.
    In the top panel, the orange box highlights the six molecular clouds studied in this work. The supermassive black hole at the center of the Milky Way, Sgr A*, is marked with a cyan asterisk. The field of view of the CSO-SHARC2 350 ${\rm \mu m}$ is outlined with gray dot-dashed lines.
    The RMS levels of the two images are approximately $\sim$0.8 Jy beam$^{-1}$ (450 ${\rm \mu m}$) and $\sim$0.3 Jy beam$^{-1}$ (850 ${\rm \mu m}$), though it is important to note that due to the stacking of multiple maps, the RMS level is not completely uniform across the field of view.
    Beam sizes are shown as black circles in the lower-left corner of each panel.
    }
    \label{fg_JCMTcmz}
\end{figure*}

Previous observational studies on some molecular clouds have found that their column density probability distribution functions (N-PDFs) can be approximately decomposed into a log-normal distribution in the low column density region and a upper-truncated power-law in the high column density region, which are joined at a critical column density $N_{\rm threshold}$ (e.g., \citealt{Schneider2015MNRAS.453L..41S,Lin2016ApJ...828...32L,Lin2017ApJ...840...22L,Chen2018ApJ...859..162C,Jiao2022RAA....22g5016J}, etc).
Theoretical studies suggested that the log-normal component traces the gas structures that are supported by turbulence or other physical mechanisms (\citealt{Vazquez-Semadeni1994ApJ...423..681V,Padoan1997ApJ...474..730P,Federrath2008ApJ...688L..79F,Federrath2010A&A...512A..81F}), while the power-law component traces the gas structures that are bound by self-gravity; the transition from turbulent (or other support mechanisms) to gravity dominant regimes occur at the column density $N_{\rm threshold}$, which depends on the energetics and may vary from cloud to cloud (\citealt{Klessen2000ApJ...535..887K}).

\begin{table*}[htb!]
    \caption{Properties of parental molecular clouds and the most massive cores within them.}
    {\hspace{-0.5cm}
    \begin{tabular}{lccccccc}
        \hline
        \hline
        \multicolumn{3}{c}{Parental molecular clouds} & \multicolumn{5}{c}{Most massive cores} \\
        \cline{1-3} \cline{4-8}
        Name & $N_{\rm cut}$ / $N_{\rm threshold}$ & $M_{\rm gas}^{\rm bound}$ & RA & Dec & Size & $M_{\rm core}^{\rm max}$ & $T_{\rm core}$ \\
         & ($\times 10^{23}$ cm$^{-2}$) & ($10^3\,{\rm M_\odot}$) & (J2000) & (J2000) & (pc) & (${\rm M_\odot}$) & (K) \\
        \hline
        Sgr B2        & $0.71$ / $2.40^{+0.06}_{-0.05}$ & $970^{+190}_{-200}$ & 17:47:20.16 & -28:23:04.65 & 0.10 & $630^{+280}_{-230}$ & $430 \pm 100$ \\
        Sgr B1-off    & $0.79$ / $3.02^{+0.10}_{-0.09}$ & $17.7^{+4.1}_{-4.3}$ & 17:46:47.06 & -28:32:07.23 & 0.11 & $24.9^{+7.6}_{-7.6}$ & $177 \pm 10$ \\
        Brick         & $0.71$ / $\quad\:-\quad\quad$  & $\lesssim$2.6       & 17:46:10.63 & -28:42:17.77 & 0.10 & $7.9^{+2.9}_{-2.6}$ & $143 \pm 23$ \\
        50 km s$^{-1}$& $1.00$ / $\quad\:-\quad\quad$  & $\lesssim$1.6       & 17:45:52.01 & -28:58:53.89 & 0.09 & $6.4^{+1.9}_{-1.9}$ & $65 \pm 2$ \\
        20 km s$^{-1}$& $0.79$ / $2.51^{+0.06}_{-0.06}$ & $38.8^{+8.9}_{-8.5}$ & 17:45:37.48 & -29:03:49.87 & 0.14 & $62^{+21}_{-20}$ & $117 \pm 15$ \\
        Sgr C         & $0.79$ / $1.40^{+0.07}_{-0.07}$ & $10.9^{+2.9}_{-2.6}$ & 17:44:40.57 & -29:28:15.99 & 0.09 & $39^{+16}_{-14}$ & $177 \pm 51$ \\
        \hline
    \end{tabular}
    }
    \vspace{2mm}
    \tablecomments{
    Columns:
    (1) Name of the molecular cloud. 
    (2) Column density threshold used to define cloud boundary ($N_{\rm cut}$) and gravitationally bound gas ($N_{\rm threshold}$) (see Section~\ref{sub:Mbg}).
    (3) Total mass of the bound gas above $N_{\rm threshold}$.
    (4)--(5) Equatorial coordinates (J2000) of the most massive core identified within each cloud. 
    (6) Size of the most massive core, defined as the equivalent diameter: $D = 2 \sqrt{S/\pi}$, where $S$ is the projected area.
    (7) Mass of the most massive core, estimated from dust continuum emission at 1.3 mm, see Section~\ref{sub:Mmmc}.
    (8) The temperatures are derived from different molecular line measurements: CH$_3$CN line fitting using the same ALMA datasets as the continuum data in this work (Sgr B1-off, Sgr C, and 20 km s$^{-1}$ cloud; Appendix \ref{appendix:tem_fit}), literature-based CH$_3$CN results (Sgr B2 and the Brick; \citealt{Moller2025A&A...693A.160M,Walker2021MNRAS.503...77W}), and NH$_3$ line modeling (50 km s$^{-1}$ cloud; Section \ref{subsub:nh3} and Appendix \ref{appendix:tem_fit}).
    }
    \label{tab:sample}
\end{table*}

In a recent study, \citet{Jiao2025npdf} found that in the solar neighborhood ($d<$1 kpc) molecular clouds and a selected sample of distant high-mass star-forming molecular clouds in the Milky Way, the star-formation rates (SFRs) and the mass of the gravitationally bound gas structures ($M_{\rm gas}^{\rm bound}$) are approximately linearly correlated.
In addition, a follow-up study on the same sample of sources showed that the mass of the most massive core in each cloud ($M_{\rm core}^{\rm max}$) is tightly correlated with $M_{\rm gas}^{\rm bound}$ (\citealt{Jiao2025mmc}, see also the related studies that reported the correlation between $M_{\rm core}^{\rm max}$ and the mass of the parent gas clumps of the most massive cores, \citealt{Lin2019A&A...631A..72L,Xu2024ApJS..270....9X,Xu2024RAA....24f5011X}).
These findings have provided a new way of understanding the widely-applied star-formation ``law", the Gao-Solomon relation, which is a relation between the infrared and HCN luminosities of star formation systems near and far (\citealt{Gao2004ApJS..152...63G,Wu2005ApJ...635L.173W,Wu2010ApJS..188..313W, Jimenez-Donaire2019ApJ...880..127J}).
In addition, these results hint that the formation of cores and stars in the self-gravitationally bound gas structures is not entirely stochastic.
Instead, they might be highly self-regulated \citep{Chavez2025MNRAS.538.2989C,Yan2023A&A...670A.151Y}, and there is a possibility to understand them in an analytic sense (c.f. \citealt{Kroupa2003ApJ...598.1076K,Yan2017A&A...607A.126Y} for related discussion).

\begin{figure*}[htb!]
    \centering
        \hspace{-0.5cm}\includegraphics[width = 14.cm]{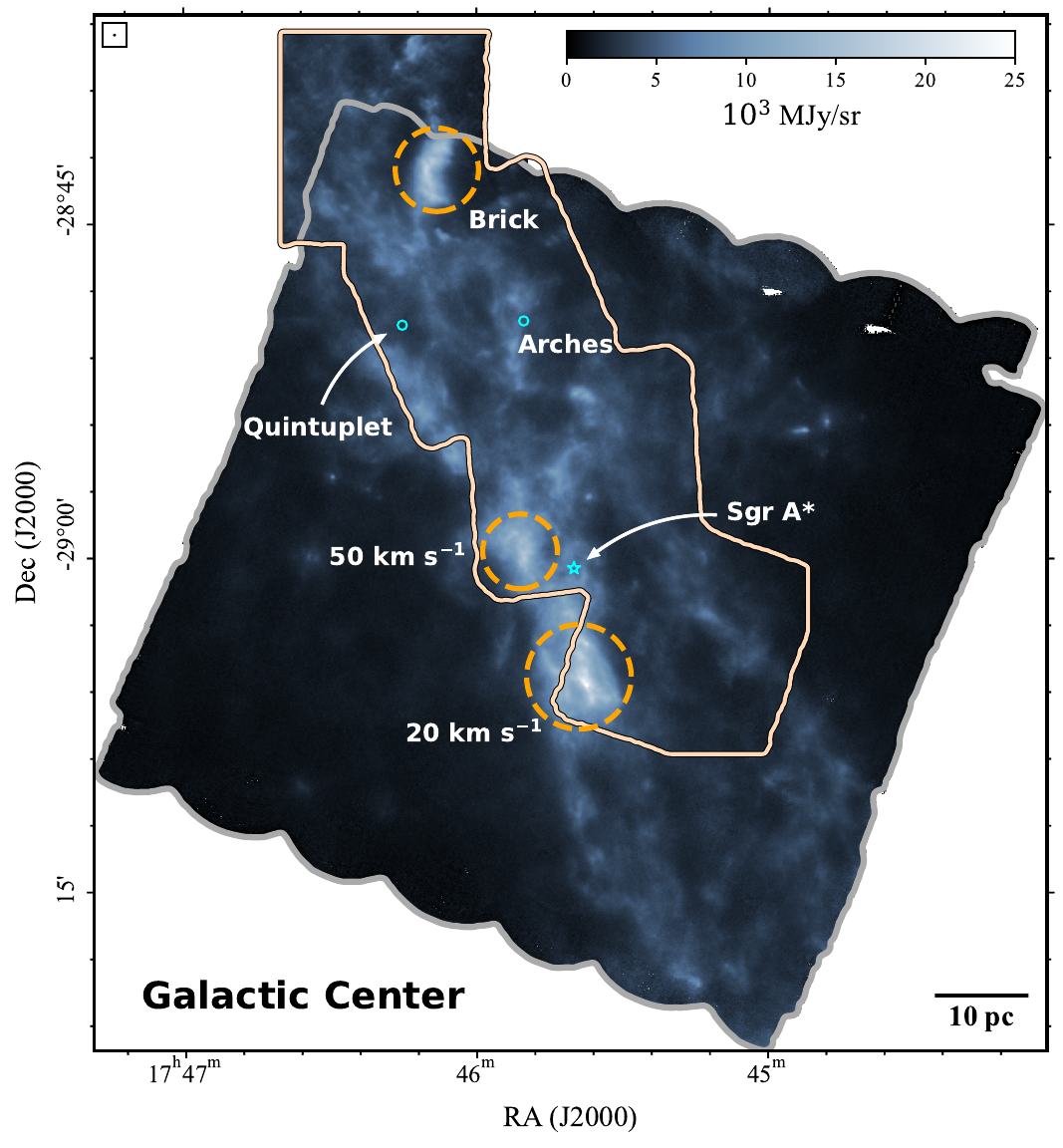}
    \caption{
    The CMZ 350 ${\rm \mu m}$ continuum map, produced by stacking the CSO-SHARC2 observations from \cite{Bally2010ApJ...721..137B_CSOCMZ} (gray outline) with our own data (light yellow outline).
    Both datasets were first combined with {\it Herschel} data before stacking. 
    Three of the selected clouds that are covered by this combined dataset are highlighted with orange dashed circles.
    The positions of two known star clusters, the Arches and the Quintuplet, are marked with cyan circles. 
    The location of the supermassive black hole, Sgr A*, is denoted by a cyan star symbol.
    }
    \label{fg_csomap}
\end{figure*}

The inner 500 pc of our Galaxy---the Central Molecular Zone (CMZ), represents one of the most extreme star-forming environments that can be studied with $<$0.01--0.1 pc resolution. 
The CMZ hosts the nearest supermassive black hole, some of the closest and most massive young star clusters, and it is the largest concentration of dense molecular gas in the Galaxy \citep{Henshaw2023ASPC..534...83H}. 
CMZ molecular clouds exhibit exceptionally high gas column densities and temperature \citep{Lis1994ApJ...424..189L,Battersby2011A&A...535A.128B,Ginsburg2016A&A...586A..50G,Tangyuping2021MNRAS.505.2392T}. They are also characterized by violent gas motions, including cloud-cloud collisions, strong turbulence \citep{Liu2013ApJ...770...44L,Minh2013ApJ...773...31M,Kauffmann2017A&A...603A..89K,Riquelme2016ApJ...824..123R}, and high velocity shear in some cases (e.g., \citealt{Liu2012ApJ...756..195L,Minh2013ApJ...773...31M}).
Specifically, the SFR of the CMZ is about ${\rm 0.09\pm0.02 M_{\odot}\:yr^{-1}}$ \citep{Barnes2017MNRAS.469.2263B}, which is one order of magnitude lower than the expectation of the Gao-Solomon relation \citep{Longmore2013MNRAS.429..987L,Kauffmann2017A&A...603A..89K,Luxing2019ApJ...872..171L}. 

In this work, our aim is to examine whether or not N-PDFs in the CMZ clouds, in general, can be decomposed into one log-normal and one power-law distribution, similar to the star-forming molecular clouds outside the CMZ (c.f. \citealt{Johnston2014A&A...568A..56J,Rathborne2014ApJ...795L..25R}).
For the clouds where the log-normal$+$power-law decomposition is appropriate, we would like to further examine whether or not the relations between their SFR, $M_{\rm core}^{\rm max}$, and $M_{\rm gas}^{\rm bound}$ are consistent with those observed outside of the CMZ.
This may help clarify the dominant physical mechanisms in the star-forming gas in this extreme environment. 

The target source selection and the observational data are outlined in Section \ref{sec:method}.
The analysis strategy is detailed in Section \ref{sec:analyses}.
The results are provided in Section \ref{subsub:MM_corr}.
The physical implication of our measurements and caveats are discussed in Section \ref{ref:discussion}.
Conclusions are summarized in Section \ref{sec:conclusion}.
Appendix \ref{appendix:50kmsTest} discusses the uncertainty in the analysis of the N-PDF of a specific selected source, the 50 km\,s$^{-1}$ molecular cloud.
Appendix \ref{appendix:combination} introduces the methodology of combining the bolometric observational images taken from ground-based and space telescopes.
Appendices \ref{Appendix:C_effects_tem} and \ref{appendix:shortspacing} discuss how the temperature assumptions and missing short-spacing (in the interferometric observations) affect the derivation of molecular core masses. 
Appendix \ref{appendix:SED_bandcoverage} demonstrates how wavelength coverage affects the accuracy of SED fittings. The data products used in this study, including the JCMT and CSO continuum maps and the SED-derived $N(\rm H_2)$, $T_{\rm dust}$, and $\beta$ maps are publicly available\footnote{\url{https://github.com/Linjing2021/CMZ_dustSED}}.

\section{Observation}
\label{sec:method}

\subsection{Source selection}\label{sub:sources}

We selected molecular clouds in the CMZ based on two main criteria.
First, we included sources with available ALMA 1.3 mm observations at resolutions of $\lesssim$0.01 pc \citep{Luxing2020ApJ...894L..14L,LiuXunchuan2024RAA....24b5009L}.
Second, we included only those sources for which SFR has been systematically investigated in previous works by \citet{Barnes2017MNRAS.469.2263B} and \citet{Luxing2019ApJ...872..171L}. 
This selection resulted in a final sample of six molecular clouds (see Table \ref{tab:sample}).
Their column densities ($\gtrsim {\rm 10^{23}cm^{-2}}$, c.f. \citealt{Battersby2011A&A...535A.128B,Luxing2019ApJ...872..171L,Tangyuping2021MNRAS.505.2392T}) are more than an order of magnitude higher than those typically observed in star-forming regions outside the CMZ (e.g., \citealt{Lin2016ApJ...828...32L,Lin2017ApJ...840...22L,Jiao2025npdf}). 

Figure \ref{fg_JCMTcmz} provides an overview of the clouds we selected.
The selected regions are either associated with ${\rm H_2O}$ masers, or harbor $\sim$0.5 pc scale gas clumps that are more massive than 10$^{3}$ $M_{\odot}$ \citep{Ginsburg2018ApJ...853..171G,Luxing2019ApJ...872..171L}.
These clouds are either actively forming high-mass stars or have the potential to form massive stellar clusters in the future.

\subsection{Data}\label{sub:data}

\subsubsection{Herschel and Planck Data}\label{subsub:herschel}
We utilized the far-infrared observations taken from the {\it Herschel} Infrared Galactic Plane Survey (HI-GAL; \citealt{Molinari2010PASP..122..314M}), which comprehensively mapped the inner Galactic plane ($|l|$ $\le$ 60$^{\circ}$, $|b|$ $\le$ 1$^{\circ}$) using the Photodetector Array Camera and Spectrometer (PACS; \citealt{Poglitsch2010A&A...518L...2P}) and the Spectral and Photometric Imaging Receiver (SPIRE; \citealt{Griffin2010A&A...518L...3G}). 
HI-GAL provides photometric imaging across five bands centered at 70, 160, 250, 350, and 500 $\mu$m, with beam sizes of approximately 8\farcs5 at 70 $\mu$m, 13\farcs5 at 160 $\mu$m, 18\farcs2 at 250 $\mu$m, 24\farcs9 at 350 $\mu$m, and 36\farcs3 at 500 $\mu$m \citep{Marsh2017MNRAS.471.2730M_HIGAL_PPMAP}.

In addition to the {\it Herschel} data, we incorporated the {\it Planck} 353 GHz continuum map from the {\it Planck} Legacy Archive \citep{Planck2020A&A...641A...6P}, which provides all-sky coverage of cold dust emission with an angular resolution of approximately 5\arcmin. The {\it Planck} data are particularly sensitive to diffuse dust emission on large spatial scales. The Planck data were primarily used to complement the JCMT 850 $\mu$m continuum data by recovering the missing large-scale flux.

\subsubsection{Caltech Submillimeter Observatory (CSO)}\label{subsub:cso}

We retrieved the Caltech Submillimeter Observatory (CSO) SHARC2  bolometer array observations at 350 $\mu$m, which have an 8\farcs5 angular resolution.
The data we retrieved includes the publicly released observations toward the inner $\sim$0.5$^{\circ}$ of the Galactic Center \citep{Bally_2017}.
The details of these observations can be found in \citet{Bally2010ApJ...721..137B_CSOCMZ}.
Since the publicly released observations did not cover the extended emission around the Brick molecular cloud, we retrieved additional SHARC2 data (the light yellow outline in Figure \ref{fg_csomap}) obtained in 2006 June as part of the Galactic Plane Survey.
These observations consisted of 13 box scans ($\sim10\arcmin\times10\arcmin$) carried out under good weather conditions ($\tau_{225{\rm GHz}}\sim0.065$), with a total on-source integration time of about 11400s.
The data were well calibrated on bright point sources and processed with the \texttt{CRUSH} software \citep{Kovacs2008SPIE.7020E..1SK}.

The publicly released SHARC2 map and the SHARC2 map retrieved by us were first stacked, and then combined with the {\it Herschel} 350 $\mu$m image using the J-comb algorithm \citep{Jiao2022SCPMA..6599511J} to mitigate the large-scale missing flux issues of the SHARC2 data.
For the detailed combination procedure, see the Appendix \ref{appendix:combination} and \cite{Jiao2022SCPMA..6599511J}.

\subsubsection{James Clerk Maxwell Telescope (JCMT)}\label{subsub:jcmt}

We retrieved the archival JCMT/SCUBA-2 450 $\mu$m and 850 $\mu$m continuum data (Proposal IDs: S22AP001, and M11AEC30), along with the public data release from \cite{Parsons2018ApJS..234...22P}.
The angular resolutions are 8$''$ and 14$''$ at 450 $\mu$m and 850 $\mu$m wavelengths, respectively (\citealt{Holland2013MNRAS.430.2513H}).
These datasets were first stacked using a weighted average with weights proportional to $1/\mathrm{rms}^2$ to improve the overall sensitivity.
We then combined the SCUBA-2 images with the images taken with the {\it Herschel} and {\it Planck} space observatories by using the J-comb algorithm \citep{Jiao2022SCPMA..6599511J} to complement the missing fluxes, which are detailed in Appendix \ref{appendix:combination}. 
The final 450 $\mu$m and 850 $\mu$m intensity maps are shown in Figure \ref{fg_JCMTcmz}.

\subsubsection{Atacama Large Millimeter/submillimeter Array (ALMA) data}
\label{subsub:alma}

We utilized the ALMA 1.3 mm continuum observations for the 20 km ${\rm ^{-1}}$, 50 km ${\rm ^{-1}}$, Sgr B1-off (also known as Dust Ridge clouds e/f), and Sgr~C molecular clouds, which were taken under the ALMA project 2016.1.00243.S (PI: Q. Zhang) and have been published as \citet{Luxing2020ApJ...894L..14L} and \citet{Xu2025A&A...697A.164X}.
These observations were carried out in the C43-3 and C43-5 array configurations, which yielded the 0\farcs25$\times$0\farcs17 synthesized beam ($\sim$2000$\times$1400 AU at 8.28 kpc). 
The fields of view are shown in Figure \ref{fg_BGcontour}.
More details of the observational setups, data calibration, and self-calibration can be seen in \citet{Luxing2020ApJ...894L..14L,Luxing2021ApJ...909..177L}.
The data were not sensitive to structures larger than 7$''$ (approximately 0.3 pc at the distance of the CMZ) due to the limit of the shortest baseline.

In addition, we utilized the 1.3 mm continuum data for Sgr~B2~main taken from the ALMA-QUARKS project \citep{LiuXunchuan2024RAA....24b5009L}. 
For our analysis, we only utilized the data taken in the C-5 array configuration, which provides $\lesssim$0.03 pc spatial resolutions. 
The largest recoverable angular scale of this dataset is $\sim$3\farcs5, corresponding to $\sim$0.15 pc at the distance of the CMZ.

Finally, we made use of a Band 6 single-pointing observation targeting the "maser core" in the Brick (G0.253+0.016), obtained with the ALMA 12-m main array as part of project 2016.1.00949.S (PI: Walker; \citealt{Walker2021MNRAS.503...77W}). 
The achieved angular resolution of $\sim$0\farcs7 corresponds to a physical scale of $\sim$0.03 pc at the distance of the CMZ, matching our resolution requirement. 
The largest recoverable scale of this observation is $\sim$6\farcs7, or $\sim$0.26 pc at CMZ.

To maintain consistency with the core identification methodology of \citet{Jiao2025mmc}, we performed imaging using the {\tt tclean} task with the {\tt uvrange} parameter set to exclude baselines shorter than 85 $k\lambda$.
This ensures the largest recoverable scale of these ALMA images of $\sim$0.11 pc, matching those of the ALMA-IMF observations \citep{Motte2022A&A...662A...8M}. 

\subsubsection{NH$_{3}$ observations taken with GBT and JVLA}\label{subsub:nh3}

We have observed the NH$_{3}$ (1,1)--(6,6) hyperfine inversion lines, the SiO 1-0 line, and the CS 1-0 line in 2011, using the NRAO Green Bank Telescope (GBT) K-band Focal Plane Array (KFPA) in (project code: GBT/11B-050; PI: H. B. Liu).
The mapping field of the NH$_{3}$ observations was approximately 16$'\times$10$'$, which covered the Galactic circumnuclear disk (CND), the 20 km\,s$^{-1}$ cloud, and the 50 km\,s$^{-1}$ cloud.
Details of the observations and data calibrations have been introduced in \citet{Liu2013ApJ...770...44L} and \citet{Minh2013ApJ...773...31M}.
The angular resolutions of the NH$_{3}$ image cubes are $\sim$30$''$. 

We utilized the Karl G. Jansky Very Large Array (JVLA) observations of the NH$_{3}$ (1,1)--(6,6) hyperfine inversion lines on the 50 km\,s$^{-1}$ cloud taken in the DnC array configuration through the project 11B-210 (PI: E. A. C. Mills).
The angular and spectral resolutions for these observations are $\sim$3$''$ and $\sim$3 km\,s$^{-1}$, respectively. 
The data calibration procedure is the same as described in \citet{Mills15}, which introduced the observations taken from the same JVLA project but for another target source. 

We used the {\tt ALMICA}\footnote{https://github.com/baobabyoo/almica} tool \citep{Liu2015ApJ...804...37L}, which is a pipeline developed based on the Miriad software package (\citealt{Sault1995ASPC...77..433S}), to combine the GBT and JVLA observations of the NH$_{3}$ lines non-linearly.
This procedure first converted the GBT image cubes to visibilities using the Miriad {\tt invert}, {\tt clean}, {\tt uvrandom}, and {\tt uvmodel} tasks.
It then jointly imaged the synthetic visibilities produced from the GBT data and the calibrated JVLA data using the Miriad {\tt invert}, {\tt clean}, and {\tt restor} tasks to produce the interim JVLA+GBT image cubes.
Finally, the procedure linearly combines the GBT image cubes with the interim JVLA+GBT image cubes to produce the final VLA+GBT image cubes, using the Miriad {\tt immerge} task.
The linear combination in the final step mitigated the issues that the {\tt clean} algorithm, in general, does not conserve flux densities, although it still induced small intensity errors due to that the Miriad {\tt immerge} task does not preserve the Gaussianity of the synthesized beam (for more discussion, see \citealt{Jiao2022SCPMA..6599511J}).
The small intensity errors induced by the Miriad {\tt immerge} task are negligible compared to the thermal noise in our present case. 
More details about the image combination will be introduced in a forthcoming paper (Mills et al. in prep).

\section{Data Analyses}
\label{sec:analyses}

\subsection{Bound Gas Mass Estimation}
\label{sub:Mbg}

\subsubsection{Dust Column Density from SEDs}
\label{subsub:Nmap}

\begin{figure*}[htb!]
    {\renewcommand{\arraystretch}{0}
    \hspace{-0.6cm}\begin{tabular}{ p{1cm} }
        \includegraphics[width = 17.5cm]{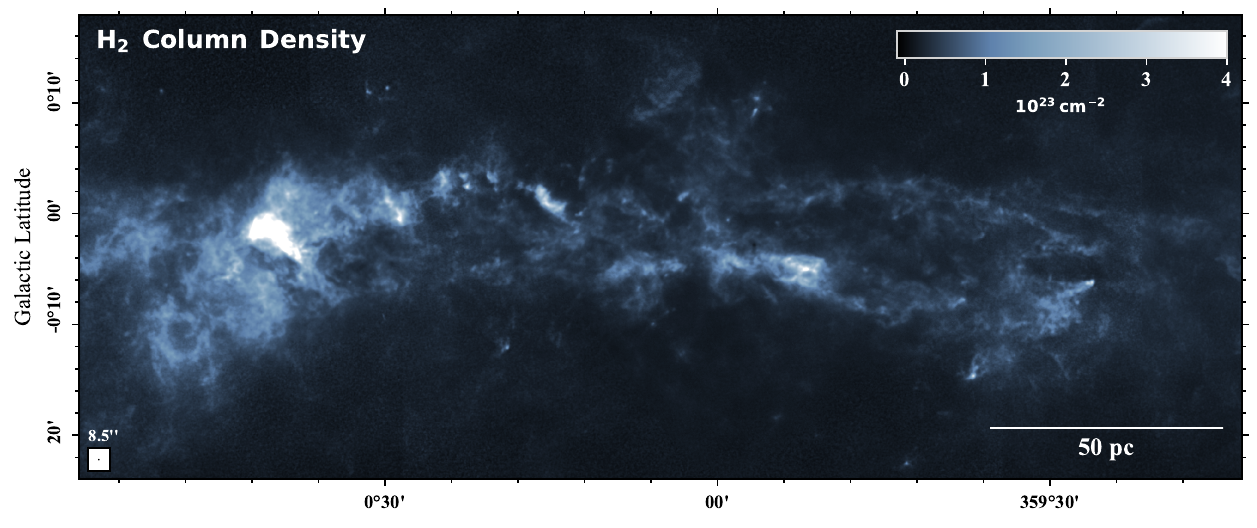} \\
        \includegraphics[width = 17.5cm]{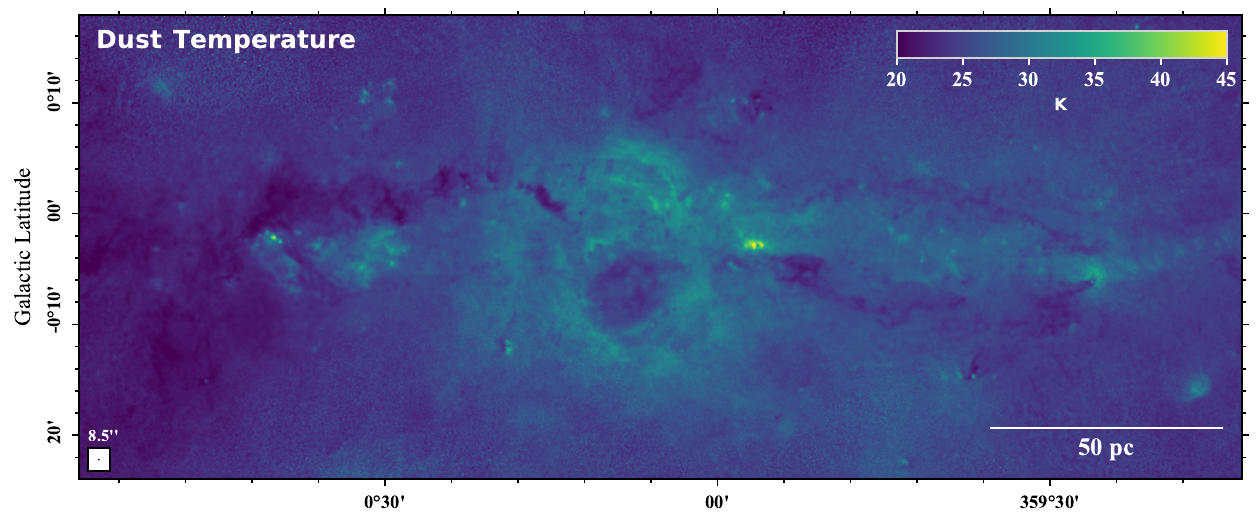} \\
        \includegraphics[width = 17.5cm]{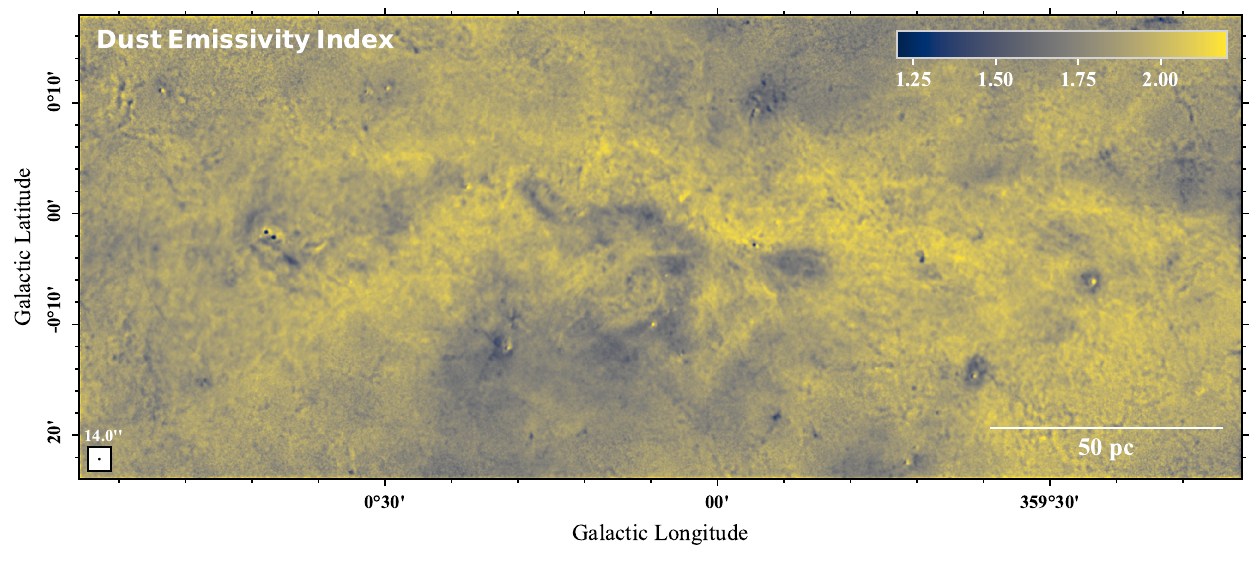} 
    \end{tabular}
    }
    \caption{
    SED fitting results.
    All maps are displayed in Galactic coordinates, with scale bars representing 50 pc. 
    Beam sizes are shown in the lower-left corner of each panel.
    {\it Top:} The 8\farcs5 resolution ${\rm H_2}$ column density ($N_{\rm H_2}$) map, derived assuming a gas-to-dust ratio of 100. 
    {\it Middle:} The 8\farcs5 resolution dust temperature ($T_{\rm dust}$) distribution map.
    {\it Bottom:} The 14\arcsec dust emissivity spectral index ($\beta$) map.
    }
    \label{fg_SEDresult}
\end{figure*}

The large-scale dust property maps---including column density, temperature, and the emissivity index---were derived by fitting the spectral energy distributions (SEDs) with a modified graybody model \citep{Hildebrand1983QJRAS..24..267H},
\begin{equation}
  S_{\nu} = \Omega B_{\nu}(T_{\rm d})(1-e^{-\tau}),
\label{eq1}
\end{equation}
where $S_{\nu}$ is the flux density at frequency $\nu$.
\begin{equation}
  B(T_{\rm d}) = \frac{{\rm 2h}\nu^3}{\rm c^2}\frac{1}{e^{{\rm h}\nu/{\rm k}{T_{\rm d}}}-1}
\label{eq2}
\end{equation}
is the Planck function for a given dust temperature $T_{\rm d}$, and $\Omega$ is the solid angle. 
The optical depth of the dust emission $\tau$ can be expressed by:
\begin{equation}
    \tau_\nu = \Sigma_{\rm dust}\kappa_\nu,
\label{eq3}
\end{equation}
where $\Sigma_{\rm dust}$ is the surface density of dust grains and $\kappa_\nu$ is the dust opacity at frequency $\nu$. The hydrogen column density $N({\rm H_2})$ can be related to dust surface density as:
\begin{equation}
    N({\rm H_2}) = \frac{\Sigma_{\rm dust} g}{\rm \mu  m_H},
\label{eq4}
\end{equation}
where, ${\rm \mu}=2.8$ is the mean molecule weight in the interstellar medium \citep{Draine2011piim.book.....D}, and $m_H$ is the mass of a hydrogen atom. 
We adopt the gas-to-dust mass ratio $g=$ 100, a commonly used value in the literature \citep{Barnes2017MNRAS.469.2263B,Battersby2025ApJ...984..156B}. While this standard assumption should be used with caution in the high-metallicity environment of the CMZ, the combined uncertainties from $g$ and the dust opacity do not significantly affect our main conclusions (see Section \ref{caveats}).
The absorption coefficient at a fixed wavelength, $\kappa_\nu$, can be related with the dust emissivity index $\beta$ as:
\begin{equation}
    \kappa_\nu = \kappa_{\rm 1000}(\frac{\nu}{\rm 1000GHz})^{\beta},
\label{eq5}
\end{equation}
where $\kappa_{\rm 1000}$ is the dust absorption coefficient at 1000GHz. 
We adopted $\kappa_{\rm 1000}=$10 cm$^2$g$^{-1}$ \citep{Hildebrand1983QJRAS..24..267H}, with 1000 GHz lying near the center of our SED fitting range (approximately 350~GHz--4~THz), making it a natural fiducial frequency that helps minimize systematic uncertainties in the dust model.

The fittings were carried out using the {\tt curve\_fit} function of the Python package \textbf{SciPy} \citep{Virtanen2020NatMe..17..261V_SciPy}, which is based on a non-linear least-squares algorithm and accounts for the noise level of each band. 
Our fittings incorporated the {\it Herschel}/PACS 70 $\mu$m and 160 $\mu$m images (Section \ref{subsub:herschel}), the CSO/SHARC2 350 $\mu$m data (Section \ref{subsub:cso}), and the 450 $\mu$m and 850 $\mu$m images that were produced by combining the JCMT/SCUBA2 images with the {\it Herschel} and {\it Planck} images (Section \ref{subsub:jcmt} and Appendix \ref{appendix:combination}).

We conducted an iterative SED fitting procedure across the CMZ following the methodology described in \cite{Lin2016ApJ...828...32L}. 
Prior to performing any SED fitting, we convolved all images to a common angular resolution matching the FHWM of the largest telescope beam at each iteration to prevent biases introduced by resolution differences.
In the first step, we used {\it Herschel} 70/160 $\mu$m (see Section \ref{subsub:herschel}), CSO 350 $\mu$m (see Section \ref{subsub:cso}), and JCMT 450/850 $\mu$m (see Section \ref{subsub:jcmt}).
All maps were convolved to match the angular resolution of the 850 $\mu$m data ($\sim$14\arcsec), which has the coarsest resolution among the input datasets. 
This step yielded large-scale maps of gas column density ${\rm N(H_2)}$, dust temperature $T_{\rm dust}$, and dust emissivity index $\beta$ at 14\arcsec resolution. 
Our multi-wavelength coverage in this step, extending from warm-dust-sensitive 70 $\mu$m emission through the graybody peak ($\sim$160 $\mu$m) to the 850 $\mu$m Rayleigh-Jeans tail, robustly constrains the parameters.
In the second step, we used the 14\arcsec map of $\beta$ obtained from the first step to initialize the SED fitting only with the 70, 350, and 450 $\mu$m images to derive dust temperature and column density maps with our best achievable angular resolution of $\sim$8\farcs5.
Here, we assume that the dust emissivity index $\beta$, which is governed by the grain size distribution and composition, remains approximately uniform within each pixel and does not vary significantly.
It is worth noting that the SHARC2 350 $\mu$m observations only cover the inner $\sim$100 pc of the Galactic Center region (see Figure \ref{fg_csomap}), limiting the inclusion of 350 $\mu$m data to this region. 
Outside the SHARC2 field of view, only 70 and 450 $\mu$m images were used in the second step.
This variation in data coverage does not affect the spatial resolution of the final SED products, but may lead to slight differences in sensitivity.

Through this two-step iterative process, we obtained a dust temperature and column density map at $\sim$8\farcs5 resolution (corresponding to $\sim$0.34 pc at 8.28 kpc, \citealt{GRAVITY2021A&A...647A..59G}), and a $\beta$ map at 14$\arcsec$ resolution ($\sim$0.55 pc at 8.28 kpc), which are presented in Figure \ref{fg_SEDresult}. For the SED results, we summarize and discuss several important caveats below:
\begin{itemize}
\setlength{\itemsep}{4pt}
\setlength{\parskip}{0pt}
\setlength{\parsep}{0pt}
    \item Including both short- and long-wavelength data provides robust SED sampling across the graybody curve, improving the reliability of the fitted parameters and mitigating the degeneracy between $T_{\rm dust}$ and $\beta$ (see Appendix \ref{appendix:SED_bandcoverage}).
    \item In a few cold, dense molecular clouds within the CMZ, the SEDs show strong mid-IR absorption features at 70~$\mu$m.
    In these regions, the derived parameters warrant extra caution due to absorption effects (see Appendix~\ref{appendix:SED_abs}). 
    The impact on column density is typically small (a few percent), but the fitted dust temperature and emissivity index can be slightly biased.
    \item We did not apply an explicit foreground/background subtraction during the SED fitting. 
    Our map ($\sim$300$\times$150~pc) is centered on the CMZ, where diffuse CMZ emission is strongly blended with large-scale Galactic cirrus, making reliable subtraction difficult. 
    Moreover, the target clouds have column densities $\gtrsim10^{23}\,{\rm cm^{-2}}$, so the $\lesssim10^{22}\,{\rm cm^{-2}}$ fore-/background contribution does not significantly affect our mass estimates or subsequent scientific conclusions. 
    However, omitting this subtraction may lead to a systematic overestimation of $\sim$ $10^{22}\,{\rm cm^{-2}}$ in column density.
    \item Mid-IR emission in Galactic-plane regions can be contaminated by non-thermal radiation from very small grains (VSGs; \citealt{Compiegne2010ApJ...724L..44C,Desert1990A&A...237..215D}), potentially elevating the 70~$\mu$m flux where the VSG fraction is high. 
    Because the VSG fraction in the CMZ is not well constrained, its impact is difficult to quantify.
    Nevertheless, in most areas our 70~$\mu$m data align well with a single graybody (see Figure~\ref{fg_SED_fitting} for an example), and the derived dust temperatures ($T_{\rm dust}\sim20\text{–}45~\mathrm{K}$) are consistent with SED results obtained without the 70~$\mu$m band \citep{Battersby2025ApJ...984..156B,Molinari2011ApJ...735L..33M}.
    These two points suggest that the contamination is likely minor, though it remains a caveat.
    \item We refer to \citet{Battersby2025ApJ...984..156B} for the publicly available, 36$''$ resolution dust column density and temperature maps that cover a larger area.
\end{itemize}


\subsubsection{Gravitationally bound gas mass}
\label{subsub:npdf}

\begin{table*}[ht!]
    \caption{Fitted N-PDF Parameters of CMZ Clouds}
        \hspace{0.5cm}
            \begin{tabular}{lccccc}
                \hline
                \hline
                Cloud & $\eta_0$ & $\sigma$ & $\eta_{\rm t}$ & $\alpha$ & $\langle N_{\rm H_2} \rangle$ ($\times 10^{23}$ cm$^{-2}$) \\
                \hline
                Sgr B2         & $-0.39 \pm 0.01$ & $0.42 \pm 0.01$ & $0.51^{+0.01}_{-0.01}$ & $-1.16 \pm 0.03$ & 1.44 \\
                Sgr B1-off     & $-0.20 \pm 0.05$  & $0.28 \pm 0.04$  & $0.14^{+0.01}_{-0.01}$ & $-3.28 \pm 0.63$ & 2.63 \\
                Brick          & $0.05 \pm 0.01$   & $0.24 \pm 0.03$  & --                      & --               & 2.11 \\
                50 km s$^{-1}$ & $-0.15 \pm 0.05$  & $0.24 \pm 0.03$  & --                      & --               & 1.63 \\
                20 km s$^{-1}$ & $-0.14 \pm 0.01$  & $0.18 \pm 0.01$  & $0.08^{+0.01}_{-0.01}$ & $-4.32 \pm 0.45$ & 2.32 \\
                Sgr C          & $-0.09 \pm 0.01$ & $0.16 \pm 0.01$ & $0.25^{+0.02}_{-0.02}$ & $-3.55 \pm 1.74$ & 1.10 \\
                \hline
            \end{tabular}
        \vspace{2mm}
        \tablecomments{
        (1) Name of the molecular cloud;  
        (2) Logarithmic peak column density $\eta_0 = \ln(N/\langle N \rangle)$ of the log-normal component;
        (3) Width $\sigma$ of the log-normal component;
        (4) Transition column density $\eta_{\rm t}$ between the log-normal and power-law components; 
        (5) Power-law slope $\alpha$ of the high-column density tail;  
        (6) Mean molecular hydrogen column density $\langle N_{\rm H_2} \rangle$.
        }
    \label{tab:npdfpara}
\end{table*}

\begin{figure*}[t]
    \hspace{-0.85cm}
    {
    \renewcommand{\arraystretch}{0.1}
    \begin{tabular}{ p{5.8cm} p{5.8cm} p{5.8cm} }
        \includegraphics[height = 6 cm]{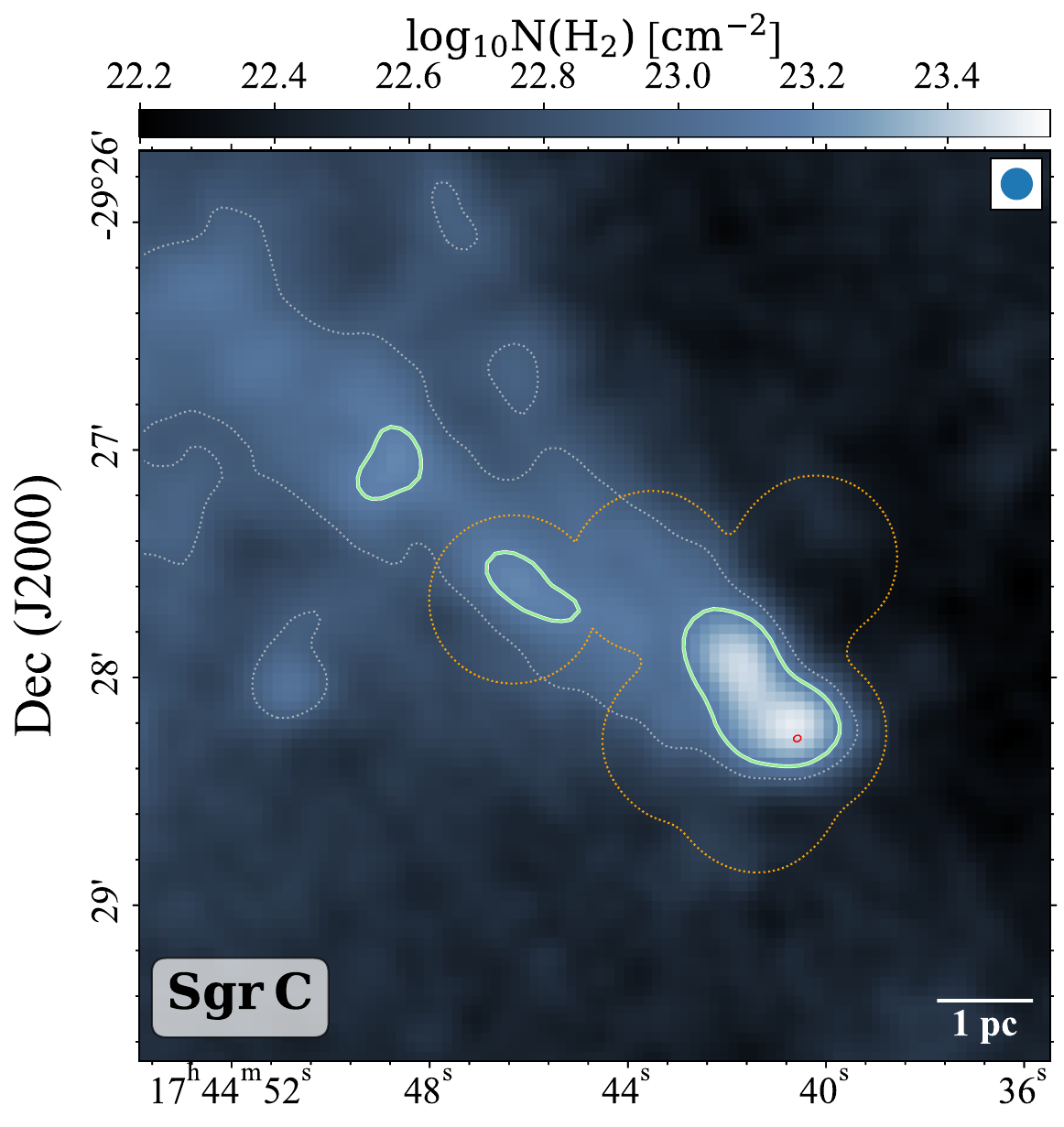} & 
        \hspace{0.25cm}\includegraphics[height = 6 cm]{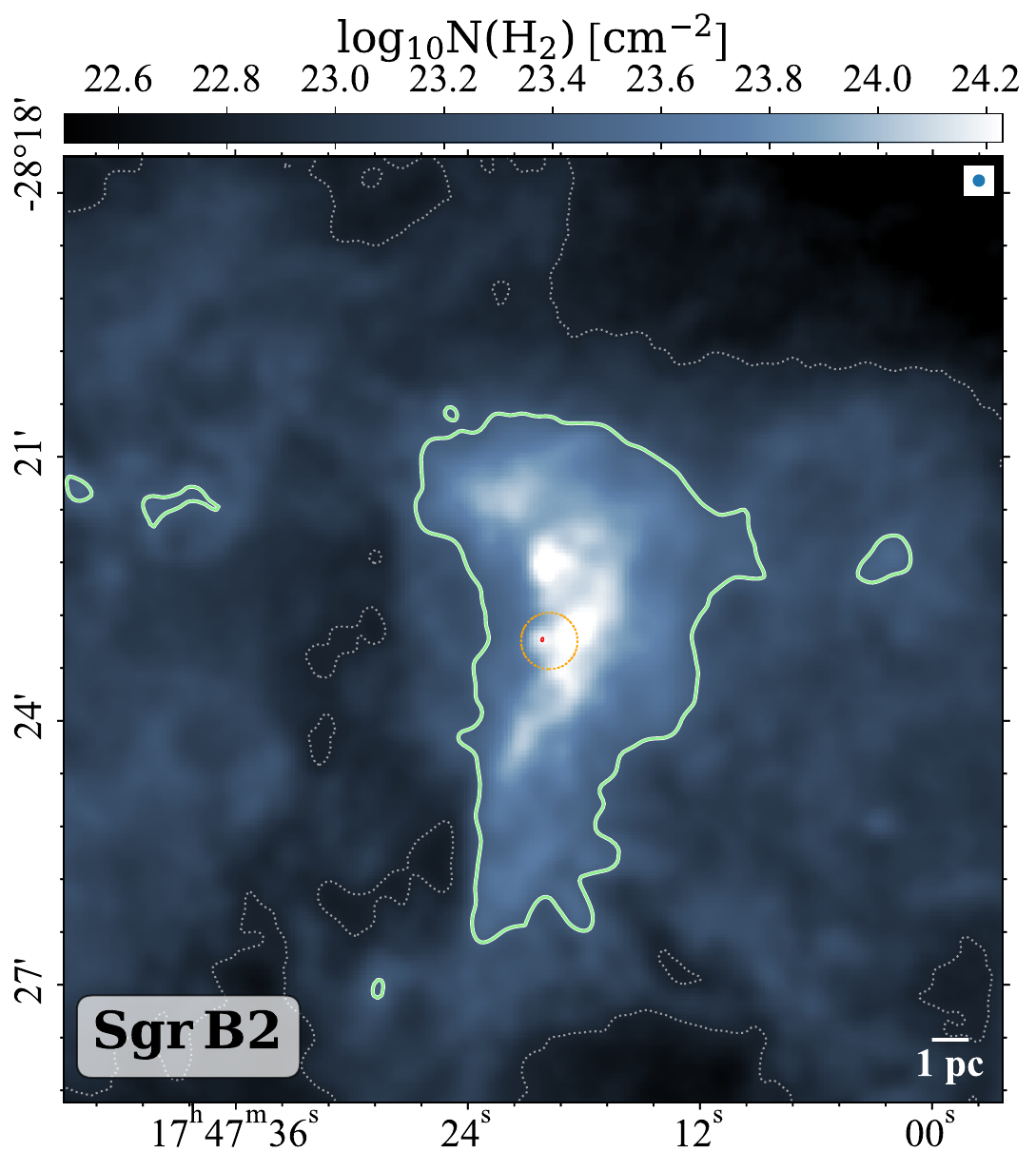} &
        \includegraphics[height = 6 cm]{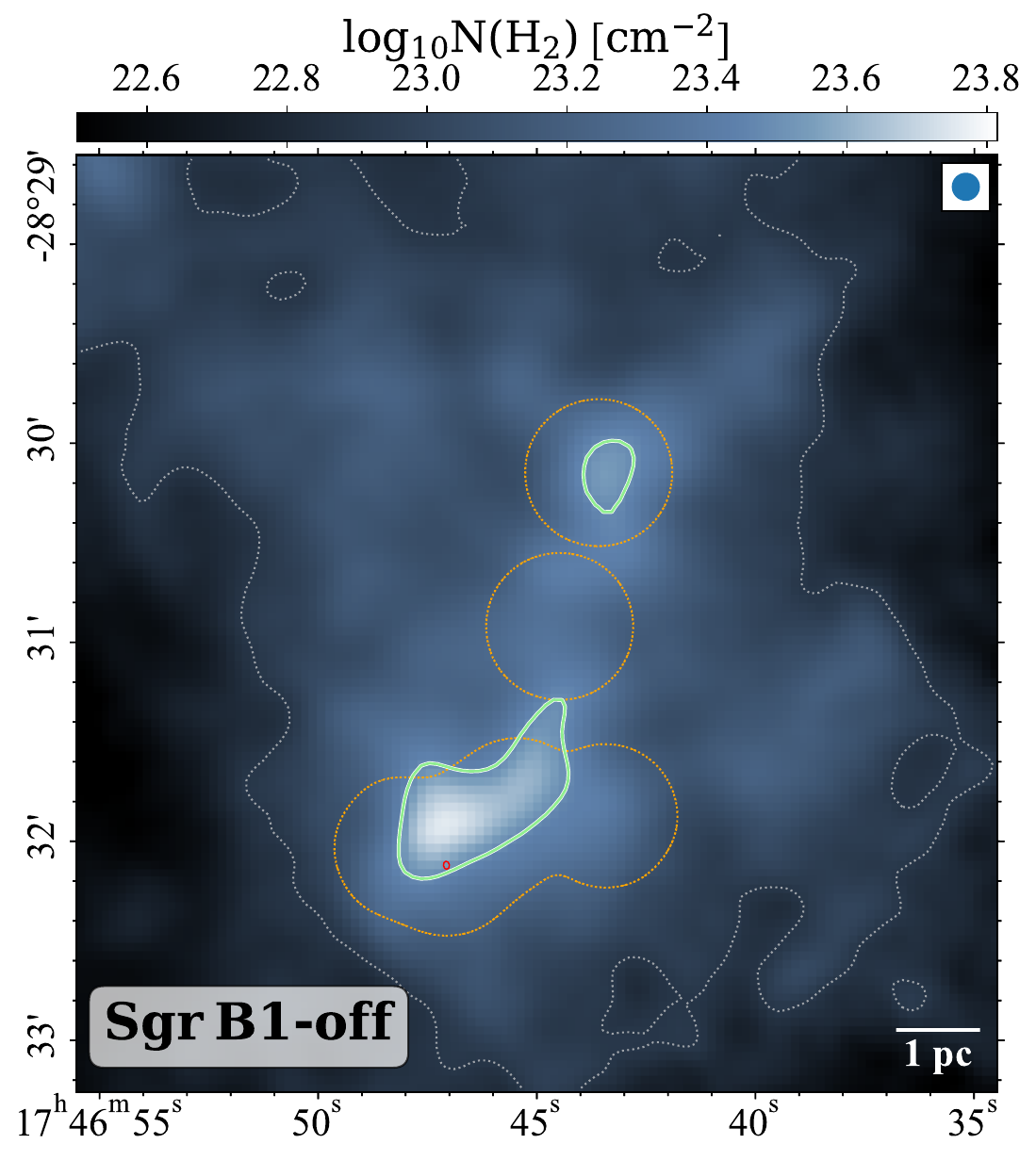} \\ 
        \includegraphics[height = 6 cm]{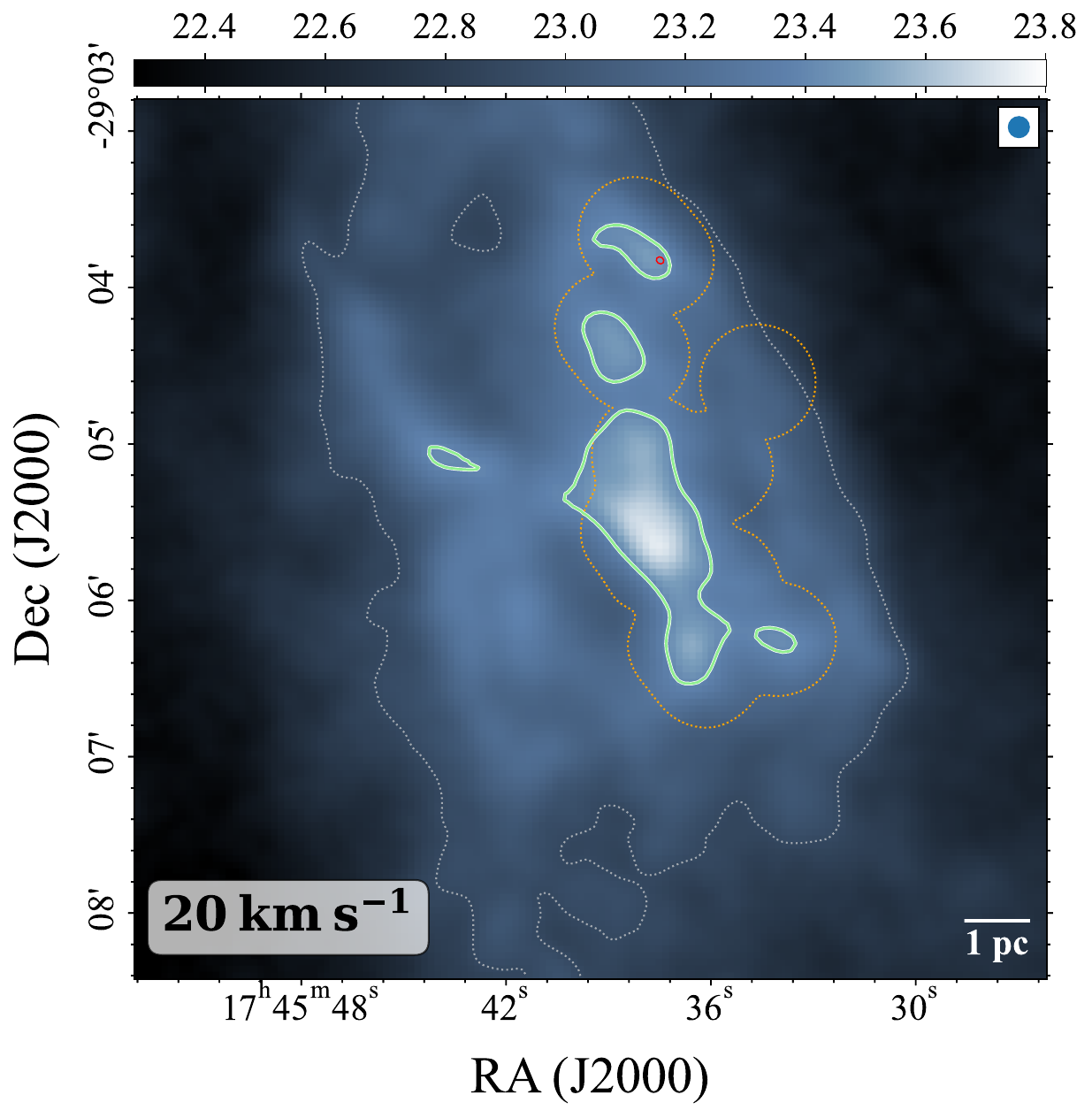}  &
        \hspace{0.25cm}\includegraphics[height = 6 cm]{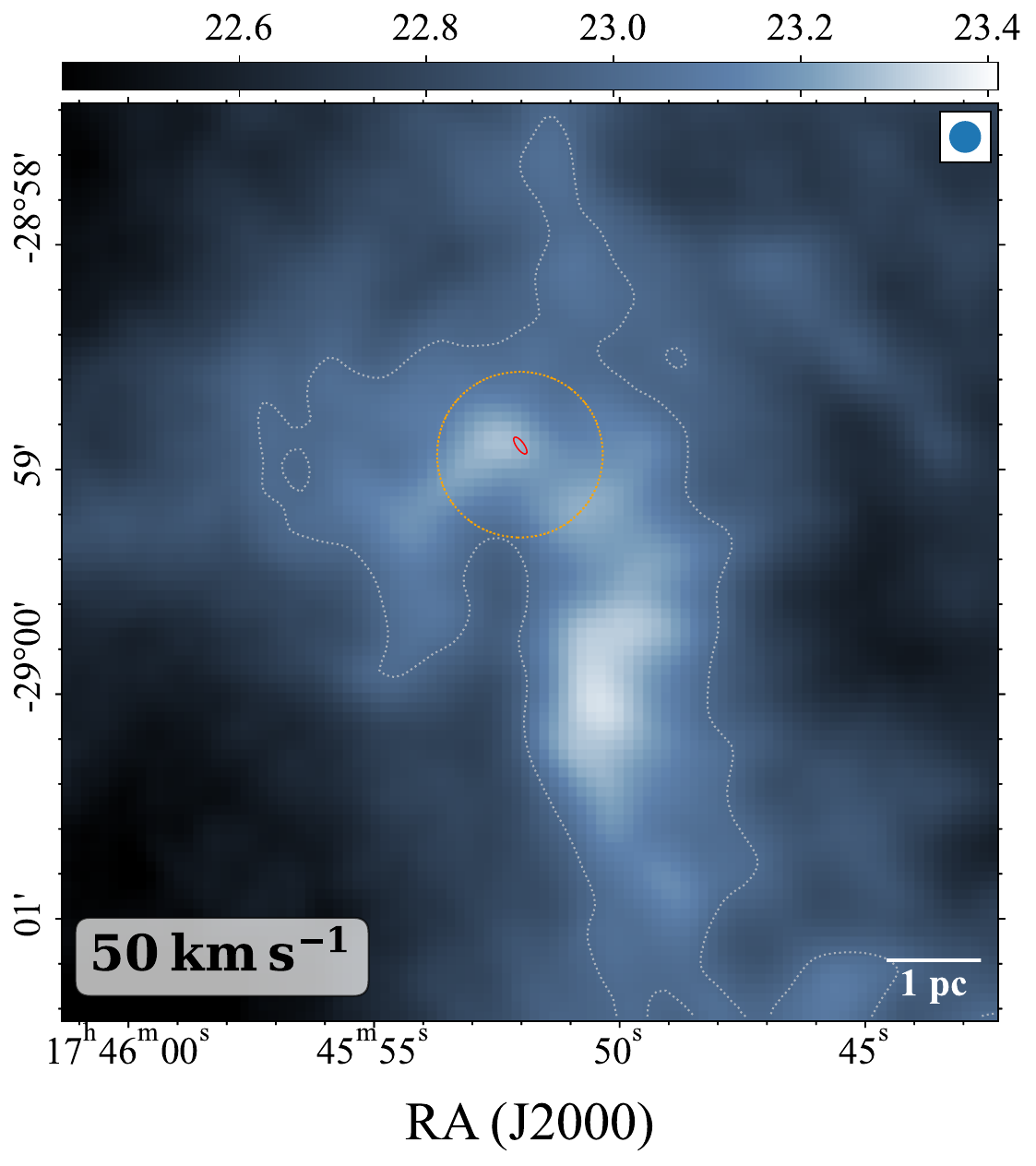} & 
        \hspace{0.07cm}\includegraphics[height = 6 cm]{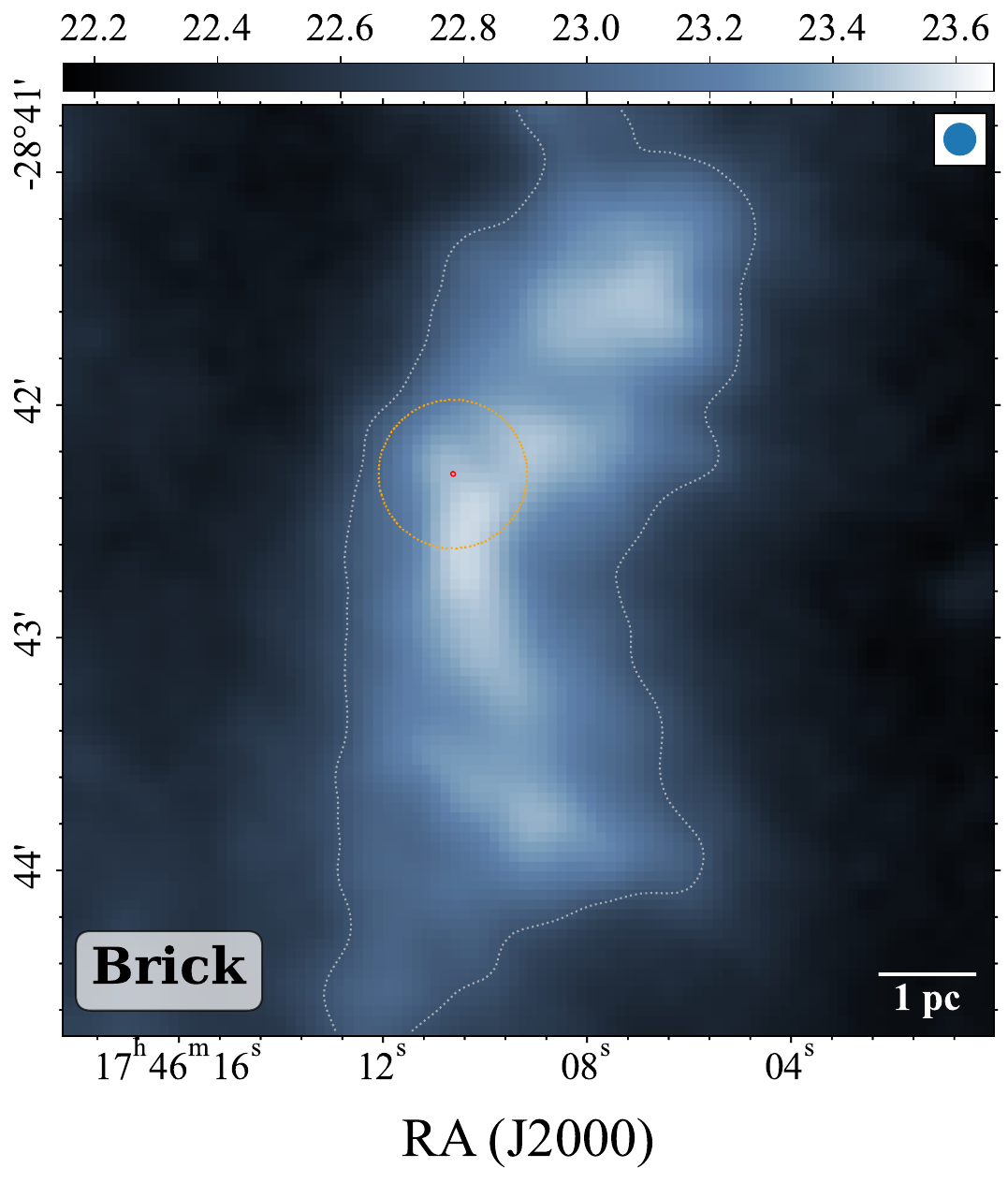} 
    \end{tabular}
    }
    \caption{The pc-scale gas column density maps of the selected sources. 
    The grayscale represents the distribution of gas column density derived from SED fitting of {\it Herschel}, JCMT, and CSO data (Section \ref{subsub:Nmap}). 
    Light green contours outline the regions of self-gravitationally bound gas. 
    The FOVs of the ALMA data are marked by the orange dotted lines, and the most massive core identified within this area is indicated by the red ellipse.
    Gray dotted lines represent the last closed contour for each source.
    }
    \label{fg_BGcontour}
\end{figure*}

\begin{figure*}[t]
    \hspace{-0.35cm}
    \begin{tabular}{ p{5.8cm} p{5.8cm} p{5.8cm} }
        \hspace{-0.95cm}\includegraphics[height = 5 cm]{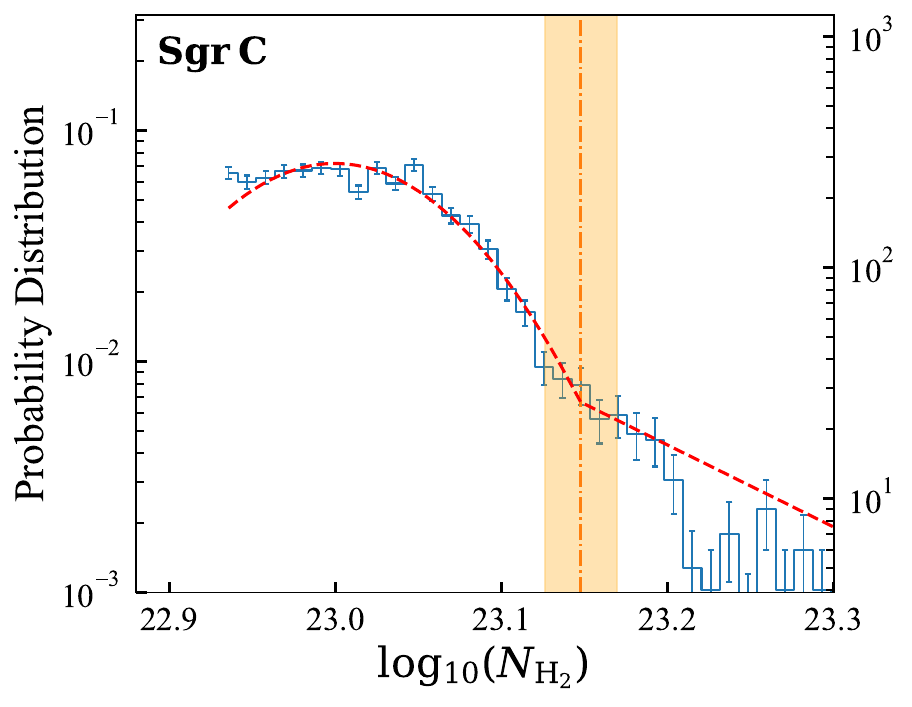} & 
        \hspace{-0.85cm}\includegraphics[height = 5 cm]{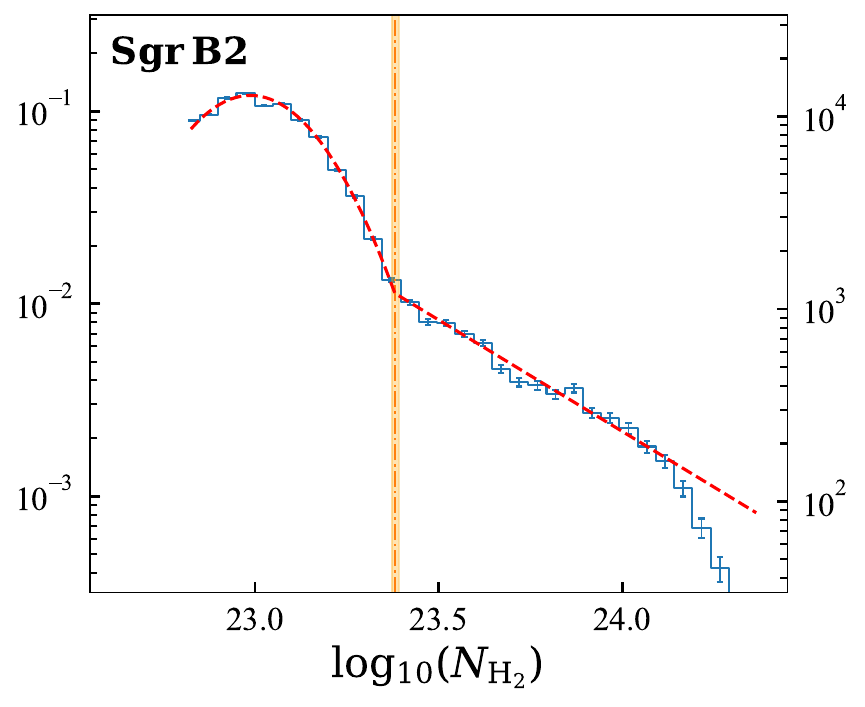}  &
        \hspace{-0.95cm}\includegraphics[height = 5 cm]{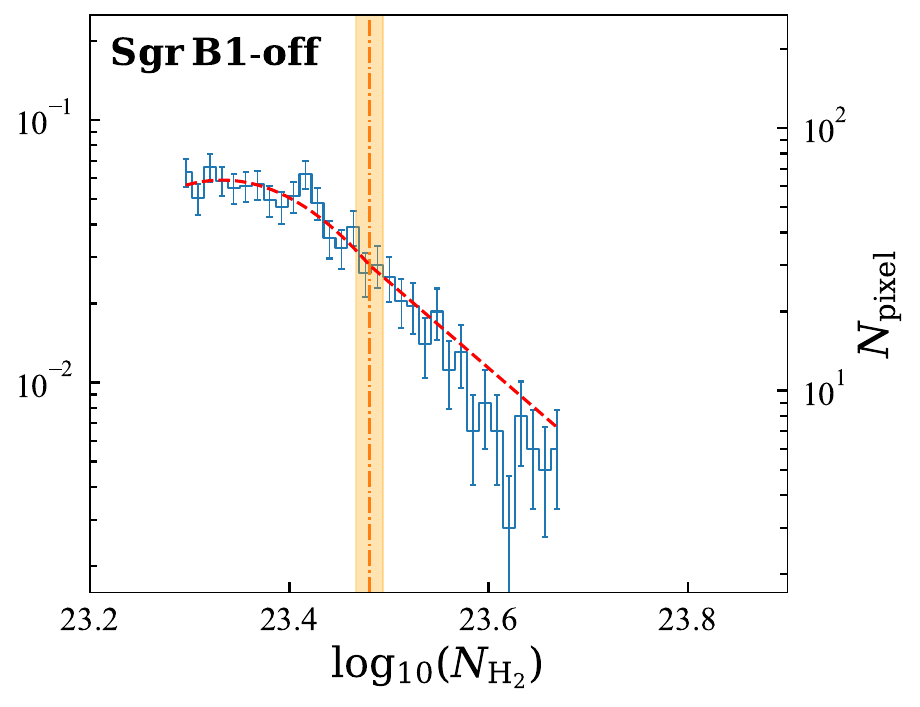} 
    \end{tabular}

    \hspace{-0.35cm}
    \begin{tabular}{ p{5.8cm} p{5.8cm} p{5.8cm} }
        \hspace{-0.95cm}\includegraphics[height = 5 cm]{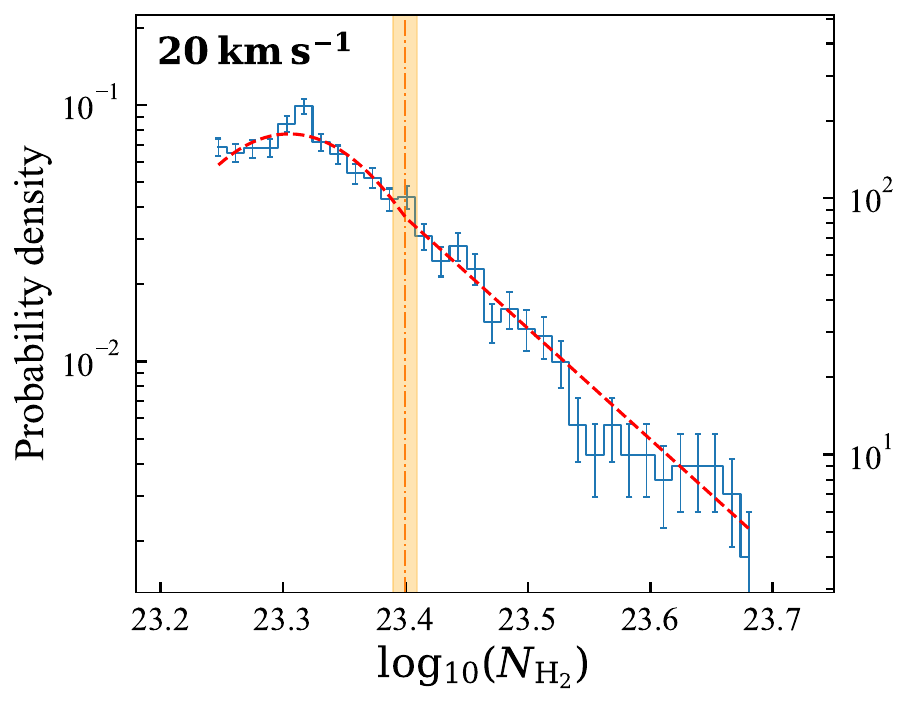}  &
        \hspace{-0.85cm}\includegraphics[height = 5 cm]{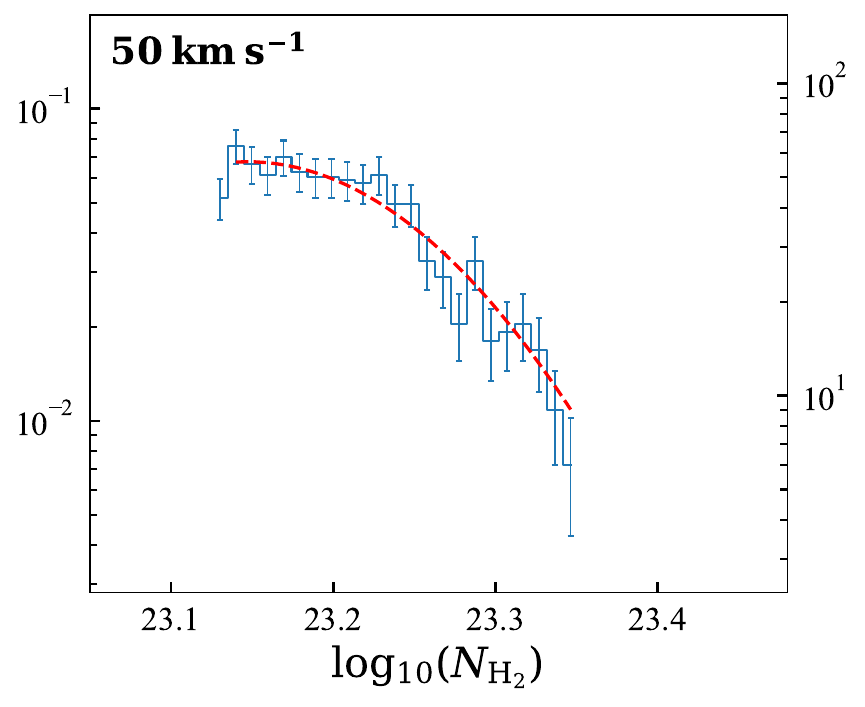} & 
        \hspace{-0.95cm}\includegraphics[height = 5 cm]{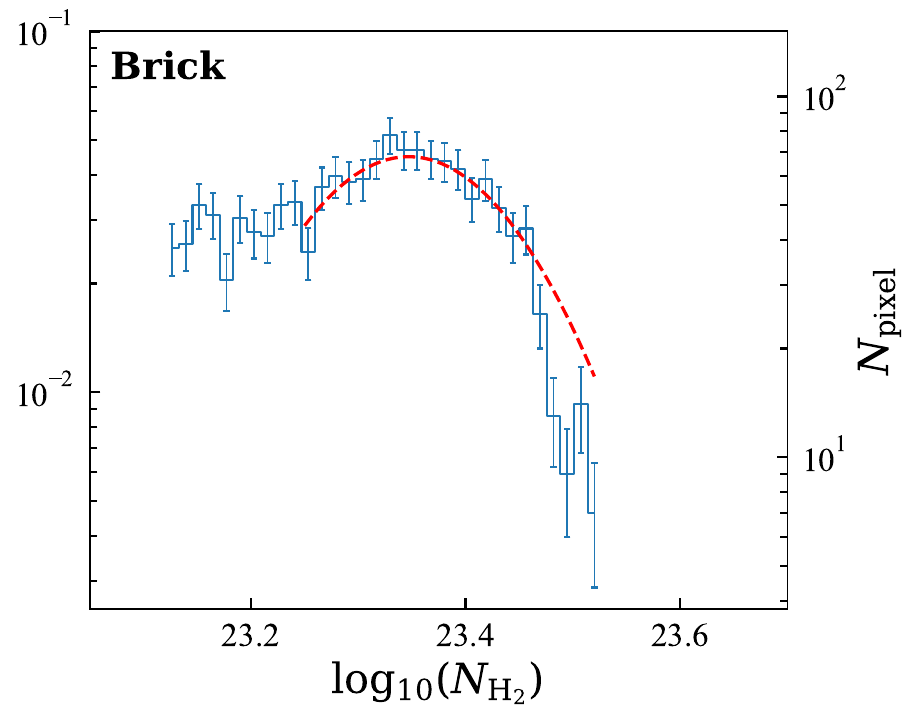}
    \end{tabular}
    \caption{
    N-PDFs of the target molecular clouds.
    The blue step lines show the N-PDFs of each molecular cloud. The left and right y-axes indicate the probability density and the corresponding number of pixels, respectively. Given our pixel size of 2\farcs5, bins with fewer than $\sim$13 pixels correspond to areas smaller than the beam size.
    The red dashed line represents the model distribution estimated using the MLE+MCMC method. The orange vertical line and shaded area represent the median of the $N_{\rm threshold}$ posterior distribution and the 3-$\sigma$ credible interval.
    The N-PDFs of the 50 km ${\rm s^{-1}}$ cloud and the Brick cannot be well described by a log-normal + power-law model, as their high column density ends do not exhibit distinguishable power-law structures. Therefore, we adopted a log-normal model alone for the fitting.
    }
\label{fg_NPDF}
\end{figure*}


We constructed N-PDFs for the six targeted molecular clouds in the CMZ, based on the gas column density maps derived from SED fittings (Section \ref{subsub:Nmap}).
We focused on column density measurements that were taken above the last closed column density contours (see \citealt{Alves2017A&A...606L...2A}) and below the high-density contours where the enclosed areas are smaller than the beam size.

When appropriate, we fit the N-PDF with the following piece-wise function,
\begin{equation}
    p(\eta) = 
    \begin{cases}
        \frac{1}{\sqrt{2\pi}\sigma}{\rm exp}[{-\frac{(\eta-\eta_0)^2}{2\sigma^2}}] \quad\quad\quad\quad\quad\quad\:\: \eta<\eta_t \\
        \frac{1}{\sqrt{2\pi}\sigma}{\rm exp}[{-\frac{(\eta_{t}-\eta_0)^2}{2\sigma^2}}+\alpha(\eta-\eta_t)] \quad \eta\geq \eta_t \\
    \end{cases},
    \label{eq:npdf}
\end{equation}
where $\eta = \ln(N_{\rm H_2}/\langle N_{\rm H_2}\rangle)$ is the logarithm of the normalized gas column density, and $p(\eta)$ is the corresponding probability density function. The parameters $\eta_0$ and $\sigma$ characterize the mean and dispersion of the log-normal component. $\eta_t$ denotes the transition point between the log-normal and power-law segments, and $\alpha$ is the slope of the power-law tail.
The procedure to fit the N-PDF is similar to what was introduced in \cite{Jiao2025npdf}.
If multiple power-law sectors present in the N-PDF (c.f. \citealt{Chen2018ApJ...859..162C}), we only fit the lowest column density one.
For sources where the power-law tails are not clearly resolved in the N-PDFs (more in Section \ref{subsub:MM_corr}), we only fit the log-normal function (i.e., the upper row of Equation \ref{eq:npdf}).

Within the Bayesian framework, we constructed a likelihood function based on Equation \ref{eq:npdf} and column densities:
\begin{equation}
    \begin{aligned}
        \mathcal{L}(\theta) = \sum_{i}{\rm log}[p_{\eta}(\eta_i|\theta)]
    \end{aligned}
    \label{eq:likelihood}
\end{equation}
where $\theta$ represents the full set of model parameters, $p_\eta$ is the probability density given by the N-PDF model (Equation~\ref{eq:npdf}), and $\eta_i$ is the value of $\eta$ at each pixel in the map.
Uniform priors were adopted for all parameters except for $\alpha$. For $\alpha$, we applied a uniform prior on $\arctan(\alpha)$ to ensure uniform sampling in slope angle.
We derived the posterior probability distributions of $\eta_{t}$ and $\alpha$ using the Markov Chain Monte Carlo (MCMC) algorithm, employing the publicly available Python package {\tt emcee} (\citealt{Foreman-Mackey2013PASP}). 
We regarded the $50^{\rm th}$, $16^{\rm th}$, and $84^{\rm th}$ percentiles of the posterior distributions as our best estimates for each parameter and the lower and upper uncertainty bounds.
An advantage of the maximum likelihood approach is that it allows us to bypass the need to bin the data prior to fitting. Instead, we directly sample the probability distribution of the data, which helps avoid the additional uncertainty that may be introduced by pre-binning the data \citep{Yogesh_2012arXiv1208.3524V_prebin_NPDF}.

We estimated the mass of gravitationally bound gas ($M_{\rm gas}^{\rm bound}$) by integrating over the regions where the gas column density exceeds the transition threshold $\eta_t$. 
The uncertainty in $M_{\rm gas}^{\rm bound}$ originates mainly from three sources: 
(1) distance uncertainties, for which we adopt a uniform 160 pc uncertainty for all CMZ clouds, based on \citealt{Reid2014ApJ...783..130R_CMZmaserDistance};
(2) the uncertainty in $\eta_t$ derived from the N-PDF fits; 
and (3) systematic measurement uncertainties, including unquantified effects from dust opacity and the gas-to-dust ratio.
For the third component, we assigned a representative uncertainty of 20\%. 
The fitted model parameters are summarized in Table \ref{tab:npdfpara}.

\subsection{Mass of the most massive cores}
\label{sub:Mmmc}

Extraction of cores in the ALMA images can be carried out using a variety of algorithms, such as \textbf{SExtractor} (\citealt{Bertin1996A&AS..117..393B}), \textbf{astrodendro} (\citealt{Rosolowsky2008ApJ...679.1338R}), etc.
This has effectively led to a method-dependent definition of `cores'.
In this work, we selected to use the \textbf{SExtractor} algorithm, which was also adopted in the previous, related study, \citet{Jiao2025mmc}.
This allows a straightforward comparison of the results without being subject to systematic biases. 
Similar to \citet{Jiao2025mmc}, before core extraction, we also smoothed the ALMA 1.3 mm continuum images to a uniform 0.03 pc spatial resolution, under the assumption that all CMZ clouds lie at the same distance of 8.28 kpc.

In the extraction process, we set the parameter \texttt{nthresh}$=$5 to exclude low signal-to-noise ratio pixels. 
The de-blending parameters were configured as \texttt{deblend\underline{~}nthresh}$=$1024 and \texttt{deblend\underline{~}cont}=$10^{-4}$ to separate two locally blended cores. 
In addition, we specified that the minimum pixels of a core must not be more than a Gaussian beam of FWHM$\sim$0.03 pc using the \texttt{min\underline{~}npix} parameter. 
The core with the highest \texttt{flux} returned by \textbf{SExtractor} is identified as the most massive core. 

After extraction of cores, we examined previous observations of ionized gas tracers to ensure that the 1.3 mm continuum emission from the identified most massive cores is primarily due to dust thermal emission (\citealt{Hildebrand1983QJRAS..24..267H}).
Specifically, we examined the VLA 22 GHz continuum data published by \citet{Gaume1995ApJ...449..663G} and \citet{Luxing2019ApJ...872..171L}.
Except for Sgr~B2~main, we find no spatial overlap between the positions of the most massive cores and any prominent 22 GHz emission source.

For the case of Sgr~B2~main, we estimated the contribution of free–free emission to the 1.3 mm flux based on the 22 GHz continuum data \citep{Gaume1995ApJ...449..663G}, assuming that the 22 GHz continuum emission is dominated by optically thin free-free emission that has a spectral index of $-$0.1.
In this estimate, in Sgr~B2~main, free-free emission contributes less than 15\% of the 1.3 mm flux density.
We might underestimate the intrinsic 1.3 mm flux density of free-free emission in Sgr~B2~main, since free-free emission in UC H\textsc{ii} regions is often optically thick, in which case the spectral index can be as high as 2 (e.g., \citealt{Keto2008ApJ...672..423K}).
Nevertheless, dust in Sgr~B2~main may be marginally optically thick (\citealt{Sanchez-Monge2017A&A...604A...6S}) and thus may attenuate the flux density of free-free emission observed at 1.3 mm wavelength, which compensates for the effect of underestimating the intrinsic flux density of free-free emission.
The uncertainty of the flux density of free-free emission will not have a significant impact on our statistical studies, as we will examine the correlation between quantities in log-space (Section \ref{sub:correlation}).

\begin{figure*}[htb!]
    \centering
        \hspace{-0.5cm}\includegraphics[scale = 0.8]{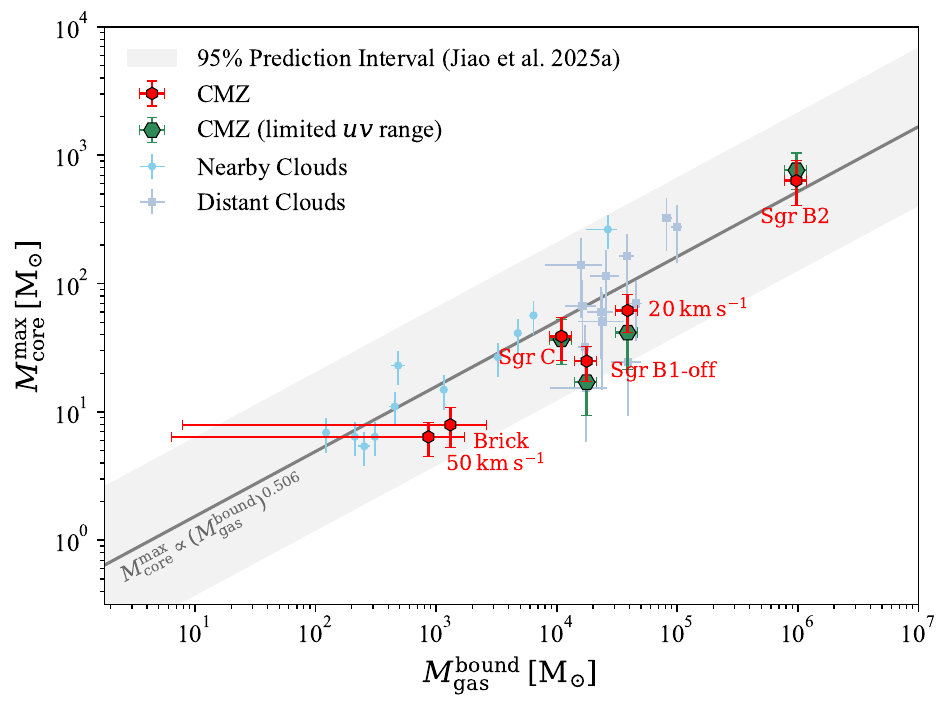}
    \caption{
    Measurements of $M_{\rm core}^{\rm max}$ and $M_{\rm gas}^{\rm bound}$ for the CMZ clouds are shown as red hexagons, and are compared with those from Solar Neighborhood clouds represented by light blue circles and distant high-mass star-forming regions shown as gray-blue squares \citep{Jiao2025mmc}. 
    The gray solid line and the light gray shaded region indicate the previously established relation and its 95\% prediction interval, respectively. 
    The green hexagons show how $M_{\rm core}^{\max}$ of CMZ clouds change after restricting the largest recoverable scale to match that of the ALMA-IMF observations \citep{Motte2022A&A...662A...8M}.
    For the 50 km s$^{-1}$ cloud and the Brick, where the resolution is insufficient to resolve bound structures, we adopt the mass within a beam at the densest region as the upper limit of $M_{\rm gas}^{\rm bound}$, and the mass of the most massive core as the lower limit. 
    We use the average of these two values as an approximate proxy for $M_{\rm gas}^{\rm bound}$ in the correlation plot. 
    }
\label{fg_MMcorr}
\end{figure*}

We used Equations \ref{eq2}--\ref{eq5} to estimate dust and gas masses in these cores ($M_{\rm core}^{\rm max}$). 
Following the method of \cite{Jiao2025mmc}, we adopt a dust opacity at 1.3 mm value of $\kappa_{1.3mm}=1$ cm$^2$\,g$^{-1}$ and assume the gas-to-dust ratio to be 100.
We assumed that gas and dust reach thermal equilibrium in the most massive cores. 
Based on the observations of ${\rm CH_3CN}$ line forests \citep{Zhang2025ApJ...980...44Z,Walker2021MNRAS.503...77W,Moller2017A&A...598A...7M} or multiple NH$_{3}$ main and satellite hyperfine inversion lines (Section \ref{subsub:nh3}), we estimated gas (and dust) temperatures assuming local thermodynamic equilibrium (LTE; more below).

We derived the temperatures for the most massive cores primarily through spectral fitting of CH$_3$CN line forests.
This molecule is an established tracer of hot core regions in massive star-forming clouds and typically traces compact structures comparable in size to the dusty cores we identified \citep{Lin2022A&A...658A.128L}. 
For Sgr B1-off, Sgr C, and the 20 km s$^{-1}$ cloud, we obtained core temperatures by fitting the CH$_3$CN line forests using the eXtended CASA Line Analysis Software Suite (\textbf{XCLASS}, \citealt{Moller2017A&A...598A...7M}) under the LTE assumption, based on data from \citet{Zhang2025ApJ...980...44Z}. Details of the fitting procedure are provided in Appendix \ref{appendix:tem_fit}.
The temperatures for the Brick and Sgr B2 main were quoted from previous ${\rm CH_3CN}$ line measurements \citep{Walker2021MNRAS.503...77W, Moller2025A&A...693A.160M}.
In the 50 km s$^{-1}$ cloud, where CH$_3$CN lines were undetected, we estimated the temperature of the most massive core using lower-resolution NH$_3$ (1,1), (2,2), (4,4), and (5,5) data.

For Sgr B2 main, the modeling included 51 transitions of CH$_3$CN with lower-state energies ranging from 58.3 to 1488 K, providing a wide $E_{\rm low}$ coverage that supports the robustness of the derived high temperature of $\sim$430 K.
We conservatively assigned a 100 K uncertainty to this measurement, reflecting reported $T_{\rm rot}$ variations exceeding 100 K among model components at individual positions \citep{Moller2025A&A...693A.160M}.
As noted by \citet{Pols2018A&A...614A.123P}, dust temperatures in Sgr B2 main may reach 900 K, significantly exceeding the CH$_3$CN excitation temperatures.
Therefore, our strategy may underestimate dust temperatures by approximately a factor of two, potentially leading to corresponding overestimates of dust and gas masses in Sgr B2 main (see Section \ref{subsub:MM_corr} for further discussion).

Table \ref{tab:sample} summarizes the derived temperatures and masses for the most massive cores. The detailed results of the spectral line fittings can be found in Appendix \ref{appendix:tem_fit} and Figure \ref{fg_Trot}.

\section{Result}\label{subsub:MM_corr}

Figure \ref{fg_BGcontour} and \ref{fg_NPDF} show the pc-scale gas column density maps (Section \ref{subsub:Nmap}) and the N-PDFs derived from them (Section \ref{subsub:npdf}) of all selected CMZ molecular clouds.
Compared to the solar neighborhood and Galactic high-mass star-forming clouds (c.f. \citealt{Jiao2025npdf}), the CMZ clouds exhibit systematically higher mean column densities.
We found that in the diffuse outskirts the column densities can exceed 10$^{22}$ cm$^{-2}$, which is consistent with what was reported in the previous studies \citep[e.g.,][]{Battersby2011A&A...535A.128B,Luxing2019ApJ...872..171L,Tangyuping2021MNRAS.505.2392T}.

The N-PDFs of the four clouds, Sgr~C, Sgr~B2, Sgr~B1-off, and the 20 km s$^{-1}$ cloud, exhibit bumps in the low column density regions, which are connected to power-law tails at higher column densities (Figure \ref{fg_NPDF}). 
We fit each of these four N-PDFs by the piece-wise (i.e., log-normal$+$power-law) function of Equation \ref{eq:npdf}, which provided the critical density $N_{\rm threshold}$ for identifying gravitationally bound gas and the value of $M_{\rm gas}^{\rm bound}$ (Section \ref{subsub:npdf}).

\begin{figure}[h!]
    \centering
        \hspace{-0.2cm}\includegraphics[width=8.5cm]{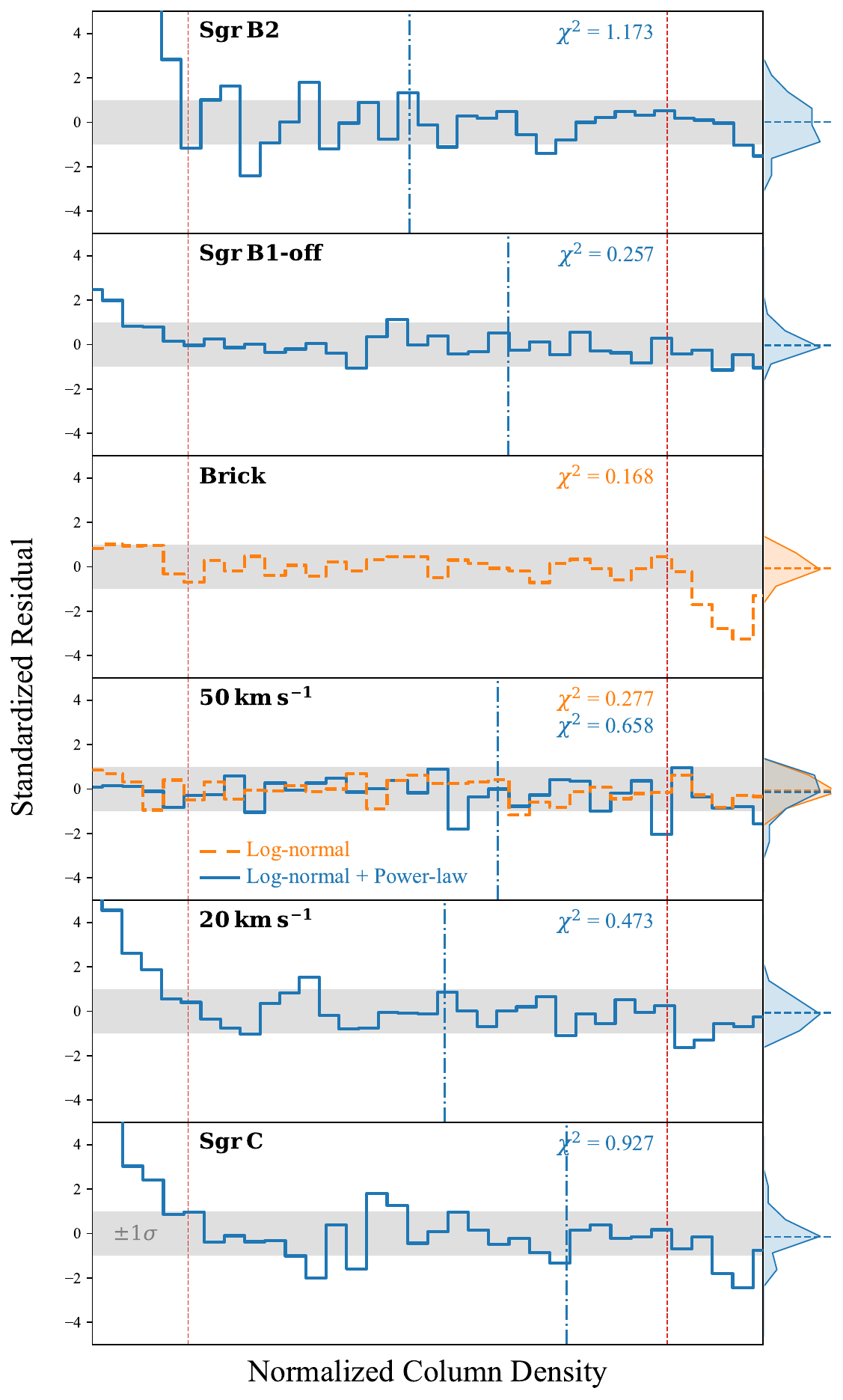}
    \caption{
    Goodness of N-PDF fitting for the six molecular clouds. Each panel shows the standardized residuals for one source, with step lines representing the difference between the observed and modeled N-PDFs, normalized by the uncertainty: $(p(\eta)_{\rm obs} - p(\eta)_{\rm model}) / \sigma$. The blue solid and orange dashed lines correspond to the residuals from the log-normal + power-law and single log-normal fittings, respectively, with the reduced chi-squared values ($\chi^2_\nu$) labeled in the upper-right corner. 
    The vertical dot–dashed lines indicate the fitted transition thresholds between the log-normal and power-law components.
    The gray shaded regions indicate the $\pm1\sigma$ range, and the red vertical dashed lines mark the fitting ranges adopted for each N-PDF. The histograms on the right display the distributions of standardized residuals, with horizontal dashed lines showing their median values. 
    }
\label{fg_NPDF_residual}
\end{figure}

For the 50 km\,s$^{-1}$ molecular cloud, the N-PDF can be fit by either a single log-normal function or a piecewise (i.e., log-normal$+$power-law) function.
When using the single log-normal, the upper limit on $M_{\rm gas}^{\rm bound}$ can be estimated as the mass enclosed within one beam area (8\farcs5, $\sim$0.34 pc) of the pc-scale column density map. 
This implies that the bound structures remain spatially unresolved at an angular resolution of 8\farcs5.
Alternatively, if the N-PDF is modeled with a log-normal plus power-law tail, $M_{\rm gas}^{\rm bound}$ is $5700^{+4400}_{-2500}$ $M_{\odot}$ (more in Appendix \ref{appendix:50kmsTest}). 
We compared the standardized residuals from both fitting approaches and found that the single log-normal fit more accurately reproduces the observed N-PDF (see Figure \ref{fg_NPDF_residual}). 
Therefore, we adopt the interpretation that the bound structures in this cloud are unresolved, and we use the mass enclosed within one beam area at the column density peak as the upper limit on the bound gas mass. 
This value is presented in Figure \ref{fg_MMcorr} and Table \ref{tab:sample}.

It is sufficient to approximate the high column density region of the N-PDF of the Brick with a single log-normal function.
The gravitationally bound gas structures in this cloud may not be spatially resolved in the pc-scale gas column density map (Figure \ref{fg_BGcontour}), which has 8\farcs5 angular resolution ($\sim$0.34 pc).
In fact, it is also not resolved in the previous ALMA mosaic observations (\citealt{Johnston2014A&A...568A..56J,Rathborne2014ApJ...795L..25R}).
For the cases of the 50 km\,s$^{-1}$ molecular cloud and the Brick, the lower limits of $M_{\rm gas}^{\rm bound}$ can be given as their $M_{\rm core}^{\rm max}$.

To evaluate how the observed N-PDFs (Figure \ref{fg_NPDF}) are represented by the piecewise (log-normal + power-law) or single log-normal models (Section \ref{subsub:npdf}), we analyzed the reduced chi-squared ($\chi^2_\nu$) statistic.
The standardized residuals are defined as:
\begin{equation}
{\rm Residual} = \frac{p(\eta)_{obs} - p(\eta)_{model}}{\sigma},
\end{equation}
where $\sigma$ denotes the Poisson uncertainty in each bin, calculated as $\sigma = N^{0.5}/N_{\rm total}$.
Here, $N$ is the number of beams in each bin and $N_{\rm total}$ is the total number of beams.

Figure \ref{fg_NPDF_residual} shows the standardized residuals for all six CMZ clouds.
The $\chi^2_\nu$ values range from 0.17 (Brick) to 1.17 (Sgr B2), with all sources satisfying $\chi^2_\nu < 1.5$, indicating overall good agreement between the models and the data.
These values are lower than typical expectations ($\chi^2_\nu \sim 1$–3), suggesting that our adopted uncertainties may be overestimated.
A possible reason is that the Poisson-based assumption of independent samples breaks down when the field of view is small and the number of independent column–density measurements is limited.
The histograms on the right of Figure \ref{fg_NPDF_residual} illustrate that the standardized residuals are symmetrically distributed around zero, indicating no systematic statistical bias between the data and the models.
Skewness tests further confirm no significant departures from symmetry for any source, except for the LN+PL fit of the 50 km s$^{-1}$ cloud, which further supports the choice of a single log-normal function as the preferred model for the 50 km s$^{-1}$ cloud.

Based on these analyses, we argue that it is appropriate to describe the N-PDFs of the molecular clouds in the CMZ using either the piece-wise (log-normal$+$power-law) function or a single log-normal function, similar to the cases of the molecular clouds outside of the CMZ (e.g., \citealt{Lin2016ApJ...828...32L,Lin2017ApJ...840...22L,Jiao2025npdf}).
Despite the extreme conditions (e.g., high temperature, high turbulence, and potentially high shear in the innermost region, c.f. \citealt{Liu2012ApJ...756..195L}) in the CMZ, we may decompose molecular clouds in it as high column density regions that are dominated by self-gravity, and regions that are dominated by other support mechanisms (e.g., turbulence, magnetic field, etc).  

Figure \ref{fg_MMcorr} shows the $M_{\rm core}^{\rm max}$ and $M_{\rm gas}^{\rm bound}$ (or the upper and lower limits) for the six selected CMZ clouds (Table \ref{tab:sample}), which are compared with the measurements from the solar neighborhood clouds (low-mass Clouds, $d<$1 kpc) and other high-mass star-forming regions (massive Clouds) reported by \citet{Jiao2025mmc}.

The four CMZ clouds, Sgr~B1-off, Sgr~B2, Sgr~C, and 20 km\,s$^{-1}$ cloud fall within the 95$\%$ prediction interval of the $M_{\rm core}^{\rm max}$-$M_{\rm gas}^{\rm bound}$ relation (Figure \ref{fg_MMcorr}) established by \citet{Jiao2025mmc}.
The p-value is 0.16 from the $\chi^2_\nu$ test, indicating no statistically significant deviation from the $M_{\rm core}^{\rm max}$-$M_{\rm gas}^{\rm bound}$ correlation established by \citet{Jiao2025mmc}. 
For the 50 km s$^{-1}$ cloud and the Brick, where the resolution is insufficient to resolve bound structures, we adopt the mass within a beam at the densest region as the upper limit of $M_{\rm gas}^{\rm bound}$, and the mass of the most massive core as the lower limit. 
We use the average of these two values as an approximate proxy for $M_{\rm gas}^{\rm bound}$ in the correlation plot, which are consistent with this $M_{\rm core}^{\rm max}$-$M_{\rm gas}^{\rm bound}$ correlation.
Here, we emphasize that this represents only a crude constraint rather than a true measurement of bound gas mass. 
Consequently, both sources are excluded from subsequent correlation analyses.


\section{Discussion}\label{ref:discussion}

\subsection{Is the $M_{\rm core}^{\rm max}$-$M_{\rm gas}^{\rm bound}$ relation universal?}\label{sub:correlation}

The consistency of the measurements from the four sources, Sgr~B1-off, Sgr~B2~main, Sgr~C, and the 20 km\,s$^{-1}$ molecular clouds (Section \ref{subsub:MM_corr}) with those taken from other star-forming regions (Figure \ref{fg_MMcorr}) indicate that the known $M_{\rm core}^{\rm max}$-$M_{\rm gas}^{\rm bound}$ relation (\citealt{Jiao2025mmc}) may be applicable at least to some sources (or certain regions) in the physical conditions as extreme as the CMZ.
How may this be understood?
Our tentative hypothesis is that, once self-gravity becomes dominant in localized regions, the core formation and star-forming activities within them are self-regulated and thus may not be sensitive to the external physical environment (e.g., turbulence, magnetic field, shear motions, external pressure due to adjacent H\textsc{ii} regions, supernovae, or cloud-cloud collisions, etc.) on scales larger than 5–10 pc \citep{Jiao2025npdf,Jiao2025mmc}.
The large-scale physical environments may play roles in determining how much material can be transported to form gravitationally bound gas structures and the timescale for the formation of gravitationally bound gas structures.

\subsection{Star Formation Rate}
\label{sub:SFR}

\begin{figure*}[ht!]
    \centering
    \includegraphics[width = 17.5 cm]{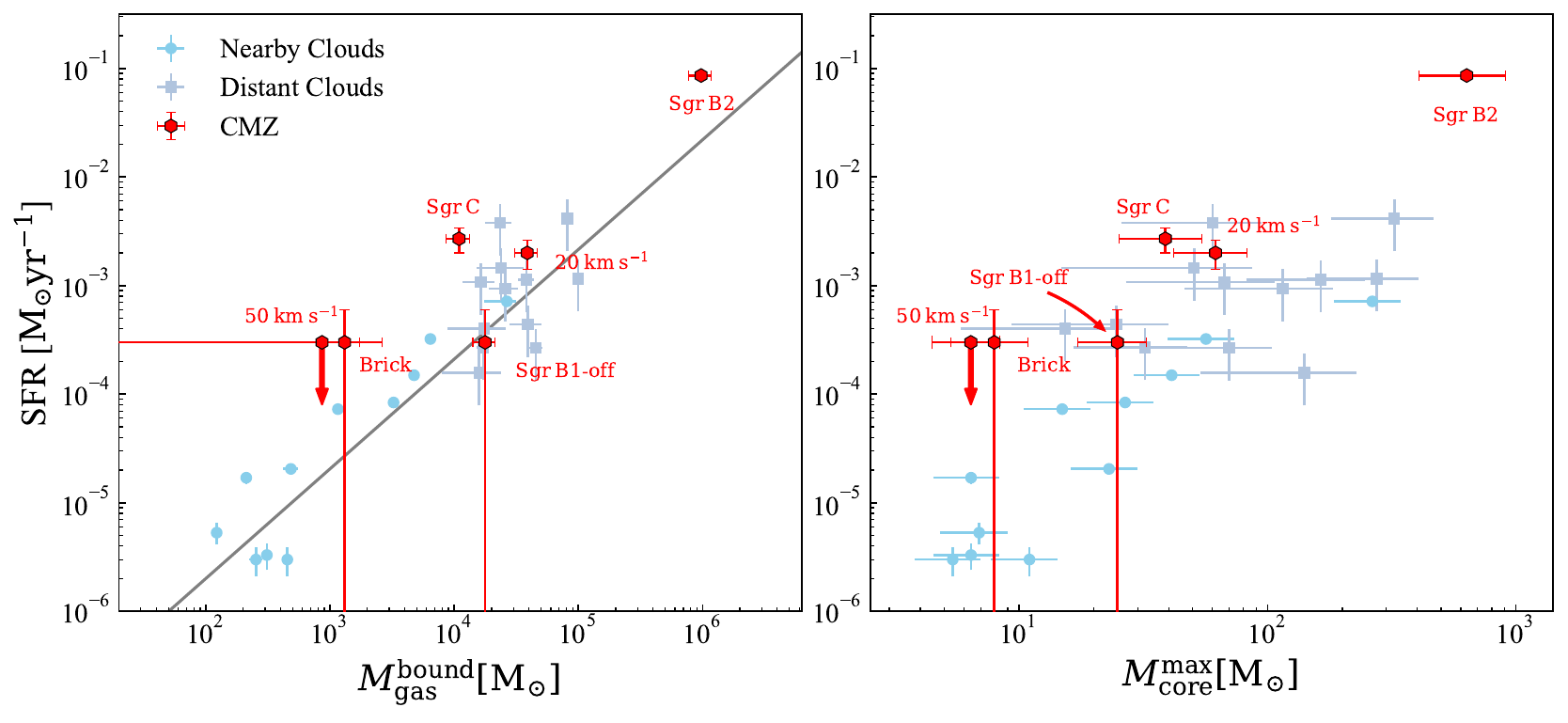}
    \caption{
    The relationships between the SFR, $M_{\rm gas}^{\rm bound}$, and $M_{\rm core}^{\rm max}$. 
    The SFRs of the CMZ clouds were quoted from \citet{Luxing2019ApJ...872..171L}, which were estimated by counting the luminous ${\rm H_2O}$ masers and the ultra-compact HII regions. 
    The $M_{\rm core}^{\rm max}$ and $M_{\rm gas}^{\rm bound}$ of the previous work are quoted from \cite{Jiao2025mmc}. 
    In the left panel, the gray solid line shows the linear relation between the SFR and the bound gas mass presented in \cite{Jiao2025npdf}.
    }
    \label{fg_SFR}
\end{figure*}

Figure \ref{fg_SFR} shows the comparison of the SFR with $M_{\rm core}^{\rm max}$ and $M_{\rm gas}^{\rm bound}$ derived from our sample (Table \ref{tab:sample}), which were compared with the measurements of \citet{Jiao2025mmc} from the star-forming regions outside the CMZ.
For the clouds in the solar neighborhood (i.e., Nearby Clouds), the SFRs were derived by counting young stellar objects (YSOs), as reported by \cite{Lada2010ApJ...724..687L} and \cite{Evans2014ApJ...782..114E}.
For distant high-mass star-forming molecular clouds outside the CMZ (i.e., Distant Clouds), the SFRs were calculated from infrared luminosities, incorporating correction factors from \citet{Jiao2025npdf} applied to original measurements by \citet{Kennicutt1998ApJ...498..541K, Gao2004ApJS..152...63G, Wu2010ApJS..188..313W}.
For our selected CMZ clouds (Table~\ref{tab:sample}), we quoted the SFRs estimated by \citet{Luxing2019ApJ...872..171L}, which were derived using counts of H$_2$O masers and ultra-compact H\textsc{ii} regions. 
In the case of the 50 km s$^{-1}$ cloud, only one weak H$_2$O maser was detected, and no signs of ongoing high-mass star formation were observed. 
Because this method is sensitive to short-timescale star formation activity, it yields only an upper limit for the star formation rate (SFR) in this cloud. 
We caution that these comparisons may be systematically biased, as the SFR estimates were derived using different tracers with potentially inconsistent calibrations (see Section \ref{caveats} for a detailed discussion).
In particular, how to convert the SFRs measured based on counting ${\rm H_2O}$ masers and ultra compact H\textsc{ii} regions to the SFRs derived from infrared luminosities remains uncertain.

Nevertheless, from the left panel of Figure \ref{fg_SFR}, we can see that among the four CMZ sources (Sgr~B1-off, Sgr~B2, Sgr~C, and the 20 km\,s$^{-1}$ cloud) in which $M_{\rm gas}^{\rm bound}$ can be constrained, SFR and $M_{\rm gas}^{\rm bound}$ are positively correlated.
The logarithmic slope is $0.92\pm 0.04$, while it is 1.01 in the Nearby and Distant Clouds samples (\citealt{Jiao2025mmc}).
The consistency of these correlation coefficients indicates that gravitationally bound gas may follow a universal star-formation law, which is our present way of understanding the well-known Gao-Solomon relation, a relation between the infrared and HCN luminosities (\citealt{Gao2004ApJS..152...63G}).
By jointly analyzing the SFRs and $M_{\rm gas}^{\rm bound}$ in the CMZ, Nearby, and Distant Clouds samples, we found the correlation:
\begin{equation}
    {\rm log}(\frac{{\rm SFR}}{\rm M_{\odot} yr^{-1}}) = (1.11\pm 0.03) {\rm log}(\frac{M_{\rm gas}^{\rm bound}}{\rm M_{\odot}}) - (7.79\pm 0.09).
    \label{eq:SFlaw_bound}
\end{equation}

Among our selected CMZ sources, $M_{\rm core}^{\rm max}$ and SFR are also positively correlated, with a power-law index of $1.59\pm 0.40$ (Figure \ref{fg_SFR}, right).
In the Nearby and Distant Clouds sample, the index is $1.79\pm 0.10$.
In the right panel of Figure \ref{fg_SFR}, Sgr~B1-off, Sgr~C, Brick, and the 20 km\,s$^{-1}$ and 50 km\,s$^{-1}$ molecular clouds seem to show a systematic upward or leftward offset from the other measurements, although similar (SFR, $M_{\rm core}^{\rm max}$) values were still observed in a few other Distant Clouds. 
This systematic offset may be attributed to the aforementioned issue of uncertain calibration between the SFRs derived based on different methods, or the time evolution of core mass or star-forming activities, which need to be clarified by future studies.

\subsection{Caveats}
\label{caveats}

While the general trends are robust, the quantitative results depend on some assumptions. 
Prominent among these are the assumed opacity law, dust-to-gas ratio, SFR measurements, and temperature equilibration between gas and dust. 
Because the last of these is discussed in Appendix \ref{Appendix:C_effects_tem}, we focus on the first three assumptions here.

Conversion of observations of dust emission to gas mass depends on the dust opacity, $\kappa_\nu$ (attenuation per surface density of dust), and the ratio of gas to dust, $g$, as shown in Equations \ref{eq3} and \ref{eq4}. 
Both can be affected by metallicity and grain growth. Commonly used opacities at long wavelengths are those of \citet{Ossenkopf1994A&A...291..943O}, usually column 5 of their Table 1 (referred to as OH5) and, more recently, \citet{Pontoppidan2024RNAAS...8...68P}, known as KP5. 
Recent analyses have found that KP5 opacities provide the best match to James Webb Space Telescope (JWST) observations of protostars (\citealt{Neufeld2024ApJ...966L..22N} and references therein).
At our fiducial frequency of 1000 GHz (300 \micron), $\kappa_\nu({\rm OH5}) = 13.6$ cm$^{2}$ g$^{-1}$ (linear interpolation in log space), while $\kappa_\nu({\rm KP5}) = 5.47$ cm$^{2}$ g$^{-1}$, larger and smaller, respectively than our value of $10.0$ cm$^{2}$ g$^{-1}$.

We have assumed that $g = 100$ in the dense regions of clouds in the CMZ. 
\citet{Giannetti2017A&A...606L..12G} found a strong dependence of $g$ on Galactocentric radius ($R_{\rm gal}$), which would imply that $g = 28$ at $R_{\rm gal} = 0$ kpc. 
However, \citet{Elia2025ApJ...980..216E} found that this dependence was too steep and favored instead a dependence that predicts $g = 38$ at $R_{\rm gal} = 0$ kpc. 
These both assume that $g = 100$ for the local ISM. 
In fact, $g$ is greater in the diffuse ISM and lower in very dense regions where most molecules other than H$_2$ have frozen onto grains by about 50\% \citep{Patra2025ApJ...983..133P}.

The likely higher metallicity in the CMZ will tend to increase  $\kappa_\nu$ and decrease $g$, both effects causing overestimation of column densities and masses of perhaps a factor of 3 to 4 using our fiducial assumptions. 
Correcting for this would move most CMZ points in Figure \ref{fg_MMcorr} closer to the fit for regions outside the CMZ, but a full analysis of the effects on all data points is beyond the scope of this paper.

Besides the gas mass measurements, we need to pay attention to the measurement of SFRs as well.
The SFR measurements used in this work are adopted from previous studies and were derived in different systems.
As described in Section \ref{sub:SFR}, the SFRs for nearby clouds were estimated by counting YSOs \citep{Lada2010ApJ...724..687L,Evans2014ApJ...782..114E}, while those for distant, high-mass star-forming molecular clouds outside the CMZ were derived from infrared luminosities \citep{Wu2010ApJS..188..313W,Jiao2025npdf}.
For the selected CMZ clouds, SFRs were estimated by counting H$_2$O masers and ultra-compact H\textsc{ii} regions \citep{Luxing2019ApJ...872..171L}.
Due to differences in timescales and sensitivity to the stellar initial mass function (IMF) across these SFR tracers, directly comparing the derived SFRs among systems is non-trivial.
For example, tracers of massive star formation, such as infrared luminosity, only reflect SFR well for clumps massive enough to have a fully sampled IMF, and old enough (5$-$10 Myr) to reach statistical steady state \citep{Krumholz2007ApJ}, while the YSO counting method for SFR focuses on small scales and shorter time scales (on the order of 1 Myr) \citep{Lada2010ApJ...724..687L,Heiderman2010,Evans2014ApJ...782..114E}.
Moreover, all SFR tracers inherently depend on the assumed IMF, which may vary across different galactic environments \citep{Yan2017A&A...607A.126Y,Jerabkova2018A&A...620A..39J,Kroupa2021arXiv211210788K,Yan2024ApJ...969...95Y,Gjergo2025arXiv250920440G}. This introduces an additional source of systematic uncertainty that remains poorly understood.
Significant discrepancies in SFR estimates derived from different systems have been reported in previous studies \citep[e.g.,][]{Lada2012ApJ...745..190L,Pokhrel2021ApJ,Elia2025ApJ...980..216E,Jiao2025npdf}.
To enable comparisons, a constant conversion factor is applied across systems \citep[e.g.,][]{Lada2012ApJ...745..190L,Jiao2025npdf}.
However, such a conversion is not yet fully justified due to the very different spatial and time scales involved, and should therefore be used with caution.


\section{Conclusion}
\label{sec:conclusion}

By jointly analyzing the existing {\it Herschel} 70, 160, 250, 350, and 500 $\mu$m images, the {\it Planck} 850 $\mu$m image, the CSO/SHARC2 350 $\mu$m images, and the JCMT/SCUBA2 450 and 850 $\mu$m images, we derived maps of the dust (and gas) column density, temperature, and emissivity index across the entire Central Molecular Zone (CMZ). The column density and temperature maps were produced at an angular resolution of 8\farcs5, while the emissivity index map has a coarser resolution of 14\arcsec. The molecular clouds analyzed in this study are representative of the more extreme regime of star-forming conditions, exhibiting high gas densities, elevated temperatures, strong turbulence, and possibly strong magnetic fields.
The reason why we selected these six specific sources was because their star-formation rates have been constrained in the recent study, \citet{Luxing2019ApJ...872..171L}, and because the high spatial resolution ($<$0.03 pc) 1.3 mm continuum images are available in the ALMA data archive.

We derived the column density probability distribution functions (N-PDFs) for these six sources, and deduced the masses of gravitationally bound gas structures ($M_{\rm gas}^{\rm bound}$) in them by fitting the piece-wise (log-normal$+$power-law) function to the N-PDFs.
We also retrieved the high spatial resolution ($<$0.01 pc) 1.3 mm continuum images for these 6 sources from the ALMA data archive, and constrained the masses of the most massive cores ($M_{\rm core}^{\rm max}$) in these regions based on the analyses of these ALMA images.

Our major finding is that the $M_{\rm core}^{\rm max}$ and $M_{\rm gas}^{\rm bound}$ are positively correlated. 
In addition, the $M_{\rm core}^{\rm max}$ and $M_{\rm gas}^{\rm bound}$ of the four sources (Sgr~B1-off, Sgr~B2, Sgr~C, and the 20 km\,s$^{-1}$ cloud) in which $M_{\rm gas}^{\rm bound}$ were successfully constrained are consistent with the known $M_{\rm core}^{\rm max}$-$M_{\rm gas}^{\rm bound}$ relation derived from a previous study of solar neighborhood clouds and some distant high-mass star-forming regions that are located outside the CMZ (\citealt{Jiao2025mmc}).
The observations of the 50 km\,s$^{-1}$ and the Brick are also consistent with this $M_{\rm core}^{\rm max}$-$M_{\rm gas}^{\rm bound}$ relation alhtough we only obtained upper and lower limits for their $M_{\rm gas}^{\rm bound}$. 
Moreover, the SFRs and $M_{\rm gas}^{\rm bound}$ of these four sources appear to follow the same correlation as what was derived from the sample of \citet{Jiao2025mmc}.
This finding led to our key hypothesis that the evolution and the star-forming activities of the gravitationally bound gas structures are insensitive to the $\gtrsim$5--10 pc scales environments exterior to them.
Instead, they may be self-regulated, although the detailed physics is yet to be investigated by studies in the future. 
Supporting mechanisms in a molecular cloud, such as turbulence and B-field, may quench the formation of self-gravitationally bound gas structures rather than quench the core and star formations in the self-gravitationally bound gas structures. 

The FITS images presented in Figures \ref{fg_JCMTcmz}, \ref{fg_csomap}, and \ref{fg_SEDresult} are available online.

\begin{acknowledgements}

This paper makes use of the following ALMA data: ADS/JAO.ALMA 2016.1.00243.S, 2016.1.00949.S, and 2021.1.00095.S.
ALMA is a partnership of ESO (representing its member states), NSF (USA), and NINS (Japan), together with NRC (Canada), MOST and ASIAA (Taiwan), and KASI (Republic of Korea), in cooperation with the Republic of Chile. The Joint ALMA Observatory is operated by ESO, AUI/NRAO, and NAOJ.
The National Radio Astronomy Observatory is a facility of the National Science Foundation operated under cooperative agreement by Associated Universities, Inc.
S.J. acknowledges support from NSFC grant nos. 12588202 and 12041302, by the National Key R\&D Program of China No. 2023YFA1608004.
H.B.L. is supported by the National Science and Technology Council (NSTC) of Taiwan (Grant Nos. 111-2112-M-110-022-MY3, 113-2112-M-110-022-MY3).
C.W.T. acknowledges support from NSFC 12588202 and 11988101.
N.J.E. acknowledges support from the Astronomy Program at the University of Texas at Austin.
X.L.\ acknowledges support from the Strategic Priority Research Program of the Chinese Academy of Sciences (CAS) Grant No.\ XDB0800300, the National Key R\&D Program of China (No.\ 2022YFA1603101), State Key Laboratory of Radio Astronomy and Technology (CAS), the National Natural Science Foundation of China (NSFC) through grant Nos.\ 12273090 and 12322305, the Natural Science Foundation of Shanghai (No.\ 23ZR1482100), and the CAS ``Light of West China'' Program No.\ xbzg-zdsys-202212.
Z.Y.Z. acknowledges the support of the National Natural Science Foundation of China (NSFC) under grants No. 12041305, 12173016,
1257030642, 12533003.

\end{acknowledgements}

\facilities{
{\it Herschel}, {\it Planck}, JCMT/SCUBA-2, CSO/SHARC2, ALMA, VLA, GBT
}

\software{{\bf Python}, {\bf CASA}
          }


\begin{appendix}

\section{Log-normal + Power-law fitting for 50 km ${\rm s^{-1}}$ cloud}
\label{appendix:50kmsTest}

\begin{figure}[h]
    \centering
    \begin{tabular} {c}
      \hspace{-0.3cm}\includegraphics[width=7.5cm]{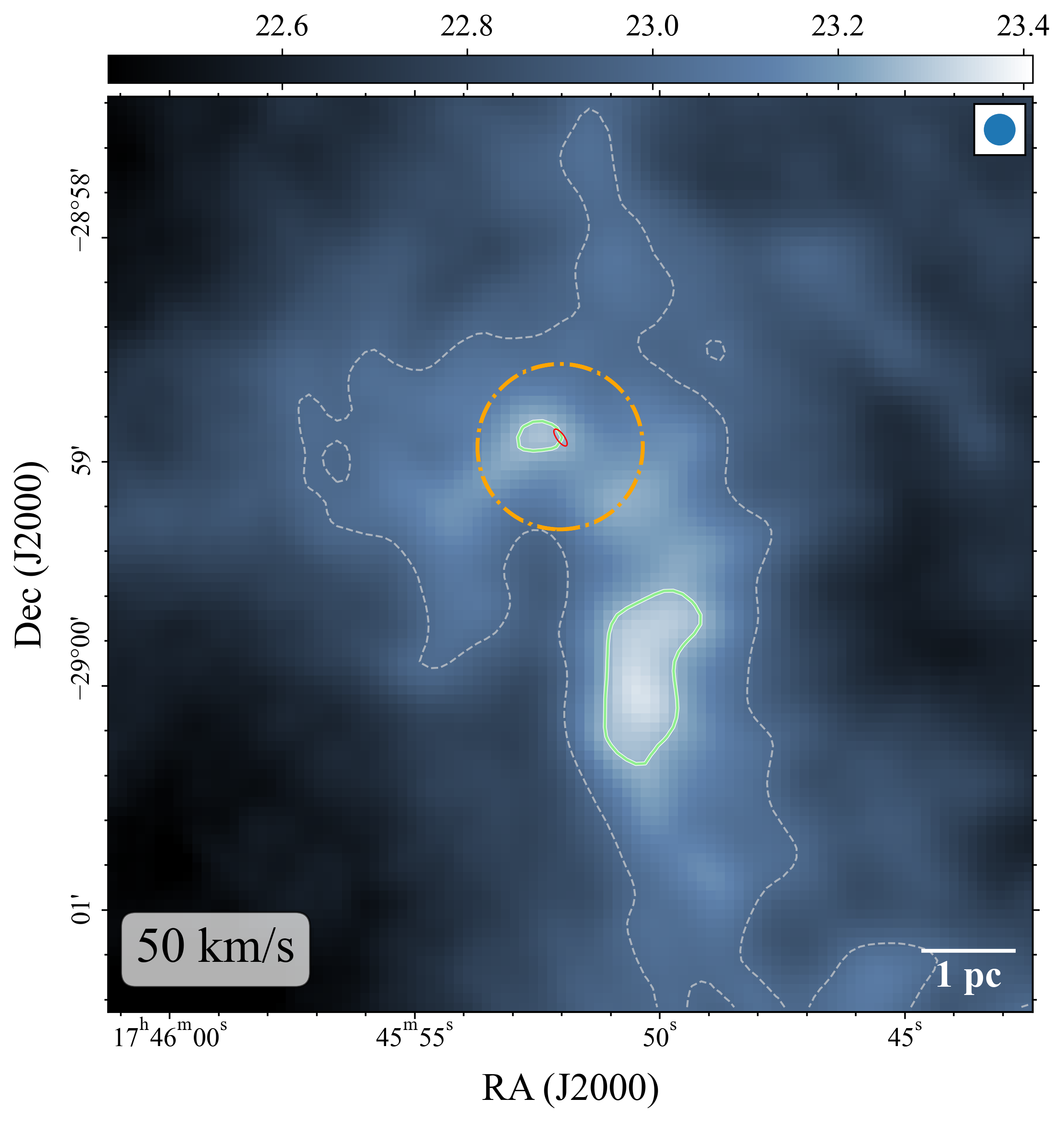} \\
      \hspace{-0.5cm}\includegraphics[width=7.5cm]{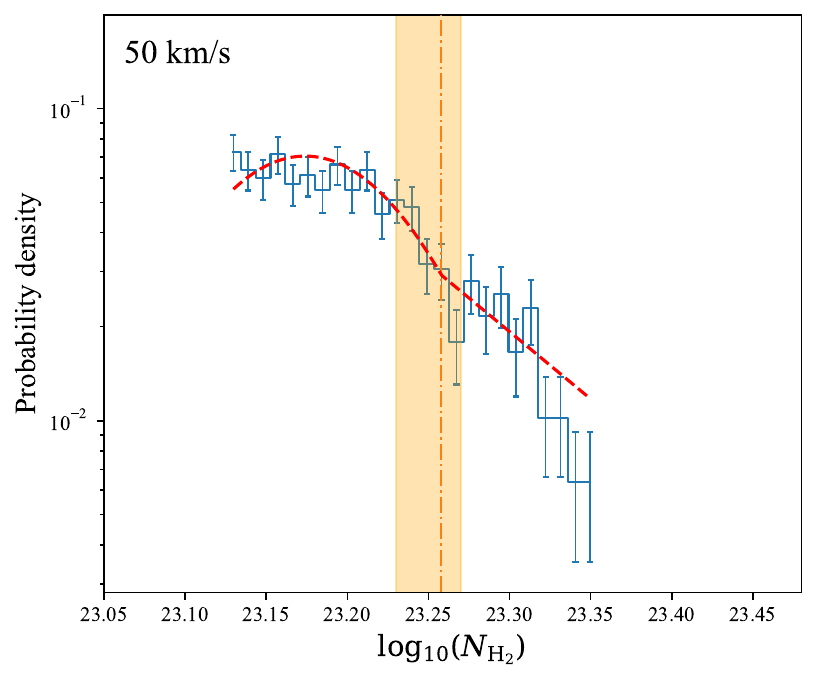} \\
      \hspace{-0.3cm}\includegraphics[width=7.5cm]{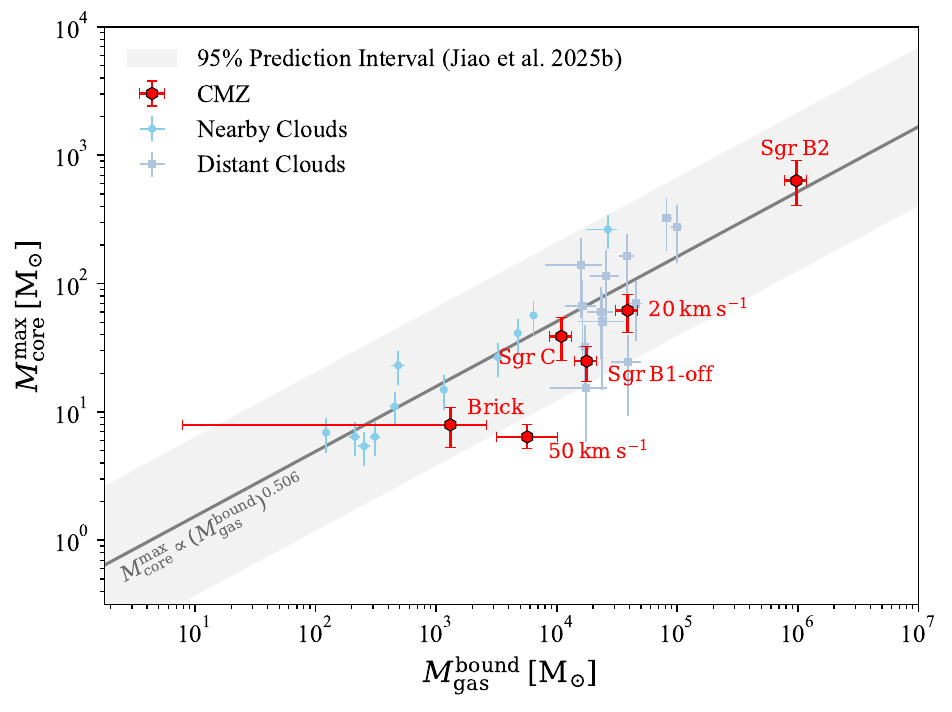} \\
    \end{tabular}
    \caption{Results of using log-normal + power-law function to fit the N-PDF of 50 km ${\rm s^{-1}}$ cloud.
    The upper, middle, and bottom panels correspond to those in Figures~\ref{fg_BGcontour}, \ref{fg_NPDF}, and \ref{fg_MMcorr}, respectively, but now applied to this alternative fit.}
    \label{fg_50kmstest}
\end{figure}

Figure~\ref{fg_50kmstest} presents our N-PDF fit for the 50 km s$^{-1}$ cloud using a log-normal + power-law function.
The fitted power-law tail spans only a narrow dynamic range in column density, and associated mass in this high-density power-law tail is $5700^{+4400}_{-2500}\:{\rm M_\odot}$.
If this component truly traces gravitationally bound gas, the 50 km s$^{-1}$ cloud would slightly deviate from the $M_{\rm core}^{\rm max}-M_{\rm core}^{\rm bound}$ relation established for molecular clouds in the solar neighborhood and Galactic disk.
Notably, while the 50 km s$^{-1}$ cloud shares comparable mass and overall density with other clouds such as the 20 km s$^{-1}$ cloud, Sgr B1-off, and Sgr C, it shows no observational evidence for embedded protostellar cores \citep{Luxing2020ApJ...894L..14L}. 
This absence of star formation activity suggests the 50 km s$^{-1}$ cloud may experience environmental conditions that suppress fragmentation or prevent the accumulation of mass on a few thousand AU scales.

If the observed power-law tail is real, it could imply that the cloud is either in an early evolutionary phase or undergoing a different fragmentation process compared to other CMZ clouds.
Follow-up high-resolution studies of its column density distribution, ideally probing down to $\sim$0.1 pc scales, will be crucial for further constraining the physical nature of its dense gas component.

\section{Combining images taken from ground based telescopes and space telescopes}\label{appendix:combination}

Bolometric (sub)millimeter continuum mapping observations obtained with ground-based facilities (e.g., CSO, JCMT, etc), in most (if not all) cases, cannot recover emission on extended angular scales.
This difficulty arises from the challenge of distinguishing the continuum emission of celestial sources from the extended thermal emission of the Earth's atmosphere. 
During data reduction, both celestial and atmospheric continuum emissions that extend beyond the simultaneous field-of-view are typically filtered out.
Without recovering the extended emission, the dust column density may be underestimated (Section \ref{subsub:Nmap}).

To address the issue of missing flux and prevent the underestimation of dust and gas column densities, \citet{Liu2015ApJ...804...37L} proposed a strategy of fusing the images taken with the ground-based and space telescopes.
This methodology was further refined through the developments of \citet{Lin2016ApJ...828...32L,Lin2017ApJ...840...22L} and \citet{Jiao2022SCPMA..6599511J}.

\begin{figure*}[ht!]
    \hspace{-1.0cm}
    \begin{tabular} { p{8cm} p{8cm} }
        \includegraphics[width=8.5cm]{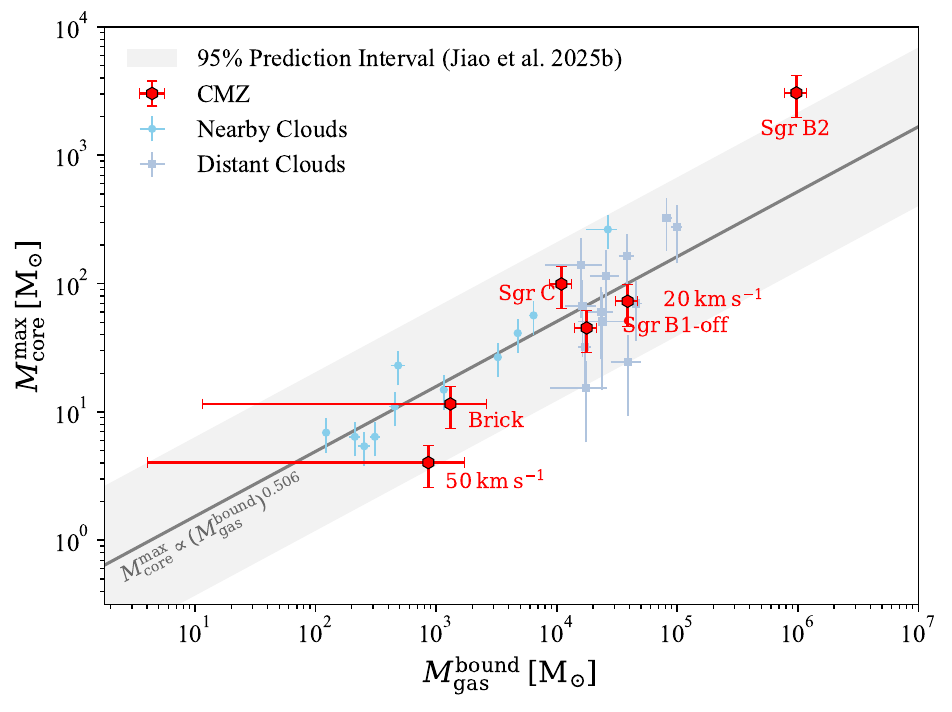} &
        \includegraphics[width=8.5cm]{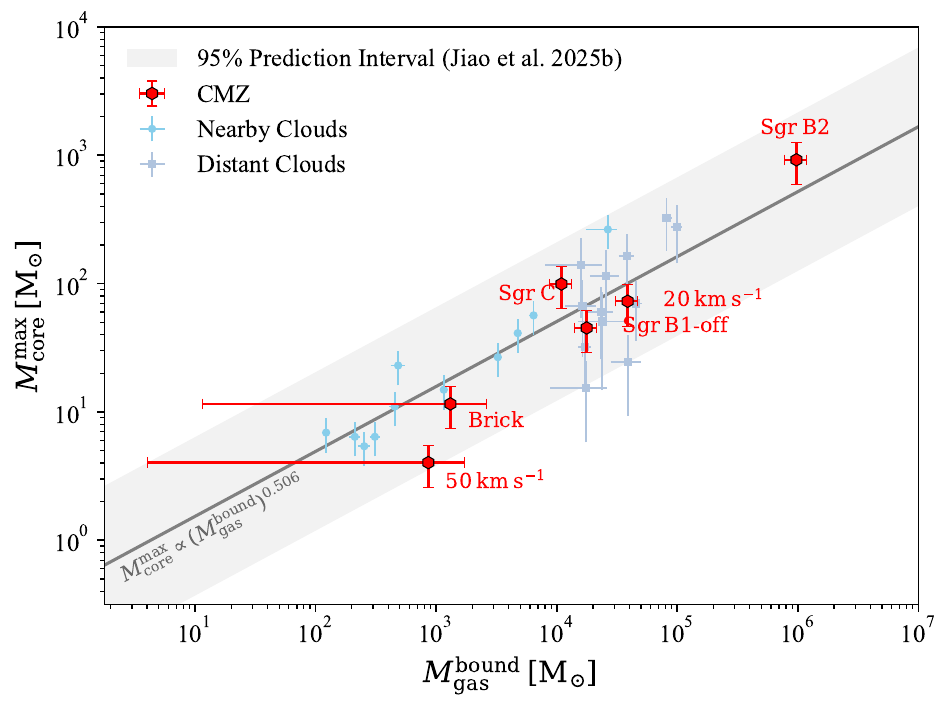} \\
    \end{tabular}
    \caption{Similar with Figure \ref{fg_MMcorr}, however, created using different assumptions of dust temperature. 
    {\it Left:--} The CMZ data points (red) were evaluated assuming a constant dust temperature of 100 K.
    {\it Right:--} The CMZ data points were evaluated assuming that the temperature in the most massive core in Sgr\,B2 is 300 K, while the temperatures in the most massive cores in the other CMZ sources are 100 K.
    }
\label{fg_MMcorr_alternative}
\end{figure*}

Our present study adopted the methodology outlined in \citet{Jiao2022SCPMA..6599511J}.
Specifically, we interpolated and extrapolated the {\it Herschel} 70--500 $\mu$m images to 450 and 850 $\mu$m wavelengths according to pixel-by-pixel fitting with the modified blackbody spectrum. 
In addition, we constructed synthetic, extended 850 $\mu$m maps by deconvolving the {\it Planck} 353 GHz map, utilizing the Lucy-Richardson algorithm \citep{Lucy1974AJ}, and employed the previously interpolated 850 $\mu$m images as priors.
We combined the CSO/SHARC2 350 $\mu$m image with the {\it Herschel} SPIRE 350 $\mu$m continuum data in the Fourier domain.
We combined the JCMT/SCUBA2 450 $\mu$m images with the previously interpolated 450 $\mu$m images and combined the JCMT/SCUBA2 850 $\mu$m images with the synthetic, extended 850 $\mu$m images mentioned earlier.

\section{Effects of temperature assumptions in the measurements of $M_{\rm core}^{\rm max}$}\label{Appendix:C_effects_tem}

A key uncertainty in estimating core masses arises from the assumption of dust temperature. 
While the typical temperature of interstellar dust is around $\sim$20 K (e.g., \citealt{Lin2016ApJ...828...32L,Lin2017ApJ...840...22L}), massive dense cores — often associated with embedded massive stars — can be significantly warmer (e.g., 10$^{2}$--10$^{3}$ K; \citealt{Wright1996ApJ...469..216W,Chen2006ApJ...639..975C,Galvan2010ApJ...725...17G,Cesaroni2011A&A...533A..73C,Liu2012ApJ...756...10L,Bonfand2024A&A...687A.163B}). 

Our present study estimated core masses based on millimeter dust thermal emission (Section \ref{sub:Mmmc}).
In this process, we used molecular excitation temperatures as proxies for dust temperatures, assuming that the observed CH$_{3}$CH and NH$_{3}$ molecules are in LTE.
 This assumption may not be unrealistic, given the high dust and gas densities in the most massive cores.
Nevertheless, to assess the robustness of the observed relationship between maximum core mass ($M_{\rm core}^{\rm max}$) and bound gas mass ($M_{\rm gas}^{\rm bound}$), we also conducted analyses based on two temperature assumptions: (1) a uniform dust temperature of 100 K for all cores, as used in \citep{Jiao2025mmc}, and (2) a two-temperature model in which we assume a temperature of 300 K for the actively star-forming Sgr B2 main core, while applying a uniform temperature of 100 K to all other cores. 
The latter assumption was motivated by the high gas temperature measured from the Sgr~B2 Main (see Table \ref{tab:sample}; c.f. \citealt{Sanchez-Monge2017A&A...604A...6S,Pols2018A&A...614A.123P}).

Figure \ref{fg_MMcorr_alternative} presents the results obtained based on the two different core temperature assumptions. 
Notably, the core masses reported in the previous work — represented by the blue and gray points in our figure — are also based on the assumption of a uniform dust temperature of 100 K for all cores \citep{Jiao2025mmc}.
Under this same assumption, our CMZ sample aligns well with the empirical correlation established in the solar neighborhood and other Galactic clouds.
In Figure \ref{fg_MMcorr_alternative}, the data point of Sgr~B2 presents an upward offset from the previously established $M_{\rm core}^{\rm max}$-$M_{\rm gas}^{\rm bound}$ relation (\citealt{Jiao2025mmc}).
Nevertheless, since the gas and dust may not be in thermal equilibrium in the most massive core in Sgr~B2~main, and the dust temperature may be as high as 900 K (\citealt{Pols2018A&A...614A.123P}), the two temperature assumptions adopted in this Appendix section underestimated the dust temperature, which in turn leads to overestimates of dust and gas masses.
Correcting this effect should make the data of Sgr~B2 move closer to the established $M_{\rm core}^{\rm max}$-$M_{\rm gas}^{\rm bound}$ relation.


\section{Effects of missing short-spacing  in the ALMA observations}\label{appendix:shortspacing}


\begin{figure}[tb!]
    \centering
        \hspace{-0.5cm}\includegraphics[width = 8.5cm]{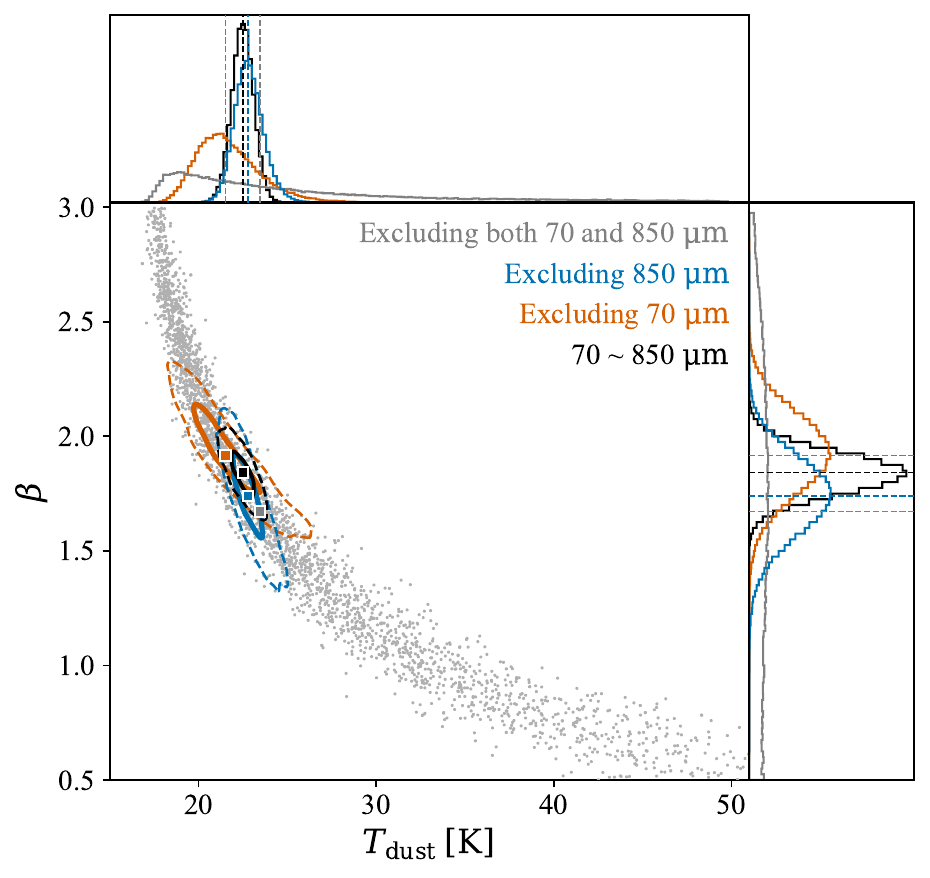}
    \caption{
    Joint posterior distributions of $T_{\rm dust}$ and $\beta$ at a representative position in the 20 km s$^{-1}$ cloud, derived from SED fitting with different band coverages.
    The black, orange, and blue contours represent fits using the full band (70-850 $\mu$m), excluding the 70 $\mu$m band, and excluding the 850 $\mu$m band, respectively. Solid and dashed contours correspond to the 0.5 and 0.1 density levels. The case excluding both 70 and 850 $\mu$m bands is shown as gray scatter points. Marginal posterior distributions of $T_{\rm dust}$ and $\beta$ are shown in the top and right panels. Colored dashed lines and square markers indicate the median values of each distribution.
    }
\label{fg_SED_corner}
\end{figure}

\begin{figure*}[htb!]
    \hspace{-1.0cm}
    \begin{tabular} { p{8cm} p{8cm} }
      \includegraphics[width=8.5cm]{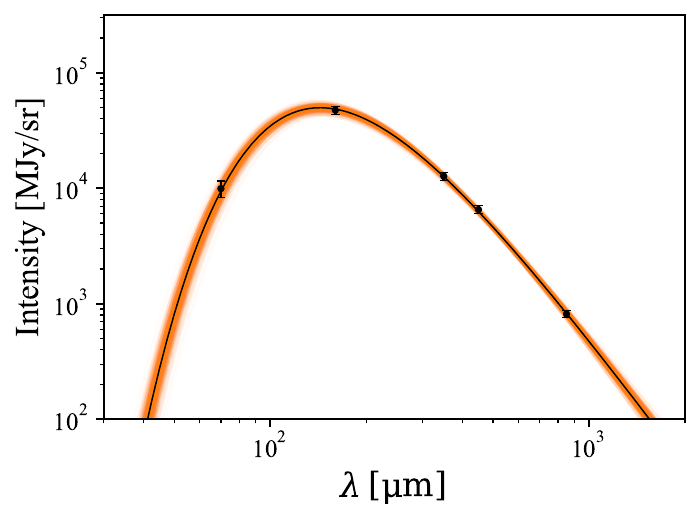} &
      \includegraphics[width=8.5cm]{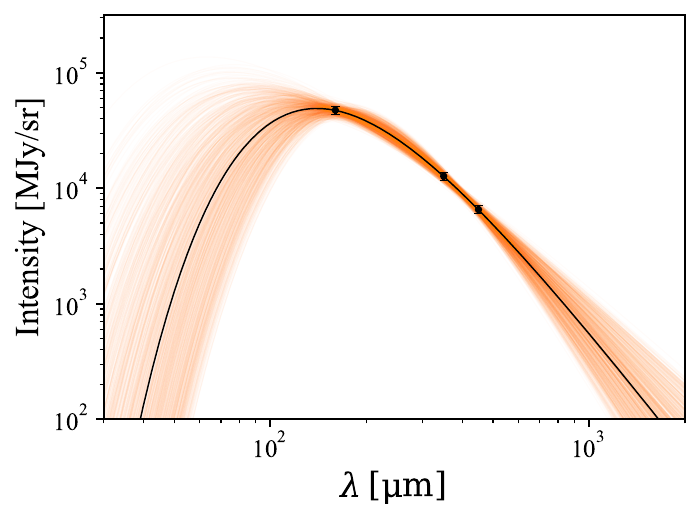} \\
    \end{tabular}
    \caption{
    Examples of dust SED fitting at the same position shown in Figure \ref{fg_SED_corner}, under two different band coverage conditions. Left: Full-band fit using 70, 160, 350, 450, and 850 $\mu$m. Right: Fit excluding both 70 and 850 $\mu$m.
    The black points with error bars indicate the observed intensities and uncertainties, while the orange curves show random draws from the posterior distribution of the modified blackbody model. The black line corresponds to the median of the posterior distribution, representing the best-estimate SED.
    }
\label{fg_SED_fitting}
\end{figure*}

\begin{figure*}[htb!]
    \centering
    \includegraphics[width=18cm]{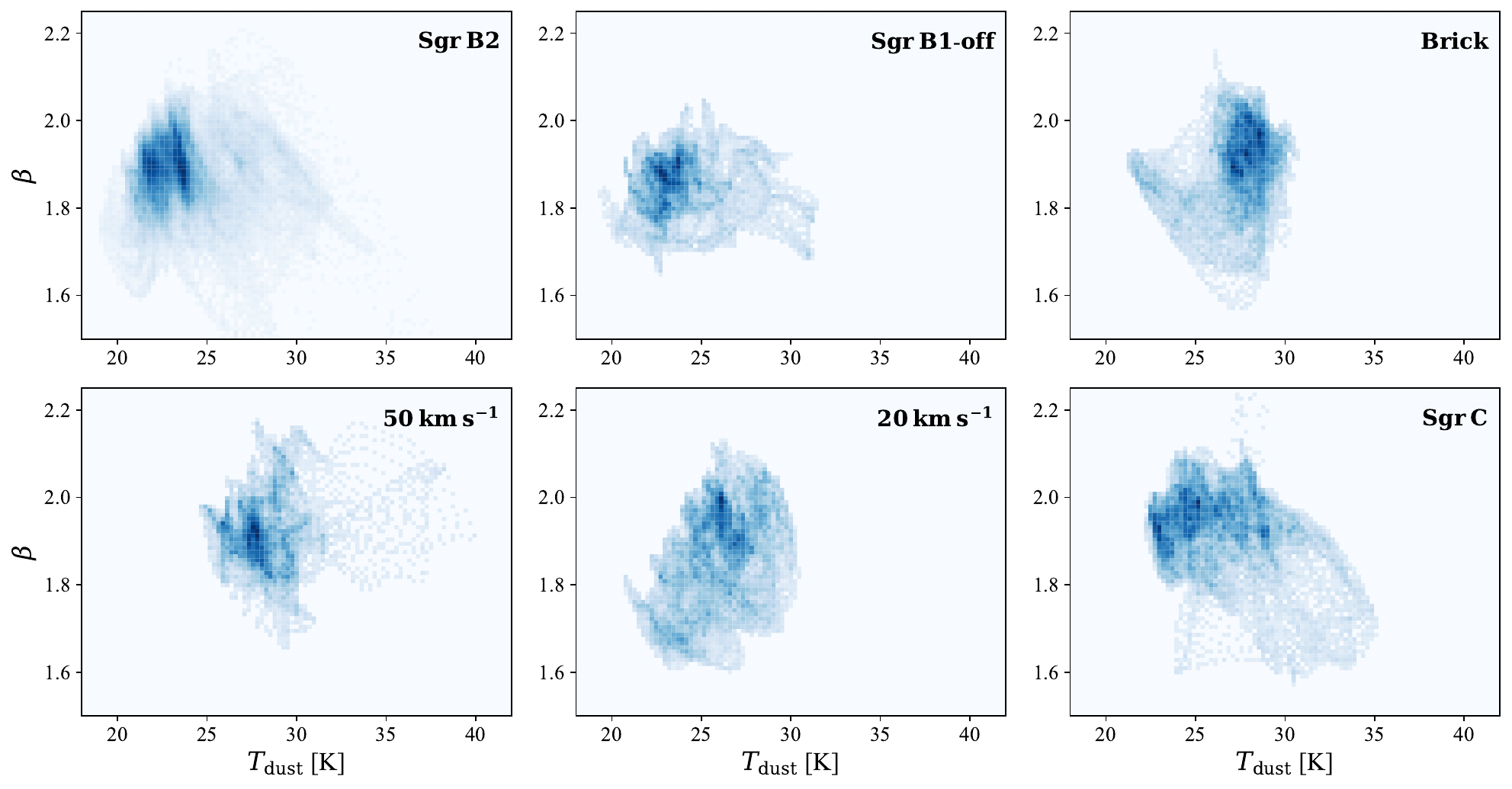}
    \caption{
    Two-dimensional histograms of $T_{\rm dust}$ and $\beta$ for the six CMZ molecular clouds studied in this work. Each panel shows the pixel-by-pixel $T_{\rm dust}$–$\beta$ distribution derived from SED fitting for one cloud.
    }
\label{fg_T_beta_distribution}
\end{figure*}

The largest recoverable scale (LRS) in ALMA observations is determined by the minimum baseline length. 
In our CMZ sample, the typical LRS of the ALMA data is approximately 0.2-0.3 pc, which exceeds the LRS of the ALMA-IMF observations analyzed by \citet{Jiao2025mmc}.
To evaluate how LRS differences affect core identification and mass measurements, we reprocessed the ALMA data for the four aforementioned sources using {\tt tclean} task from the Common Astronomy Software Applications (CASA, \citealt{CASA2022PASP..134k4501C}). 
During imaging, we limited the uvrange parameter to $>85\,k\lambda$, to match the median LRS of ALMA-IMF observations ($\sim$0.11 pc). 
To preserve the resolution required for our analysis while maximizing the recovery of the flux from the most massive cores, we adopted `natural' weighting, multiscale deconvolution, and the {\tt auto-multithresh} masking algorithm.

The resulting core masses after applying the {\it uv}-range restriction are shown as green hexagons in Figure \ref{fg_MMcorr}. 
In all targets, the change in the most massive core mass is less than 30\%, indicating that our main conclusions are robust against the differences in LRS across datasets.

\section{Factors Affecting Dust SED Fitting}

\subsection{Wavelength Coverage and the $T$-$\beta$ Degeneracy}
\label{appendix:SED_bandcoverage}

Spectral profiles of interstellar far-infrared and (sub)millimeter dust emission are consistent with the modified black body formulation, which is characterized by dust temperature ($T_{\rm dust}$), column density ($\Sigma_{\rm dust}$), dust opacity at a reference frequency, and dust emissivity index (c.f. Equations \ref{eq1}--\ref{eq5}; \citealt{Hildebrand1983QJRAS..24..267H}).
When fitting observed SEDs or spectra with the modified black body formulation, dust emissivity spectral index $\beta$ is mainly constrained by observations in the Rayleigh-Jeans limit, while $T_{\rm dust}$ is constrained by observations at shorter wavelengths.
It has been known that when the wavelength coverage is limited, the fitting will be subject to a degeneracy between $T_{\rm dust}$ and $\beta$.

To assess how the wavelength coverage impacts the robustness of SED fitting and the degeneracy between $T_\mathrm{dust}$ and $\beta$, we conduct a focused test at a representative position within the 20 km s$^{-1}$ cloud (RA = 17$^{\rm h}$45$^{\rm m}$44$^{\rm s}$.30, Dec$= -$29$^{\rm d}$05$'$01\farcs56; J2000). 
This region exhibits typical values of $N(\mathrm{H_2}) \sim 2 \times 10^{23}$ cm$^{-2}$, $T_{\rm dust} \sim 22$ K, and $\beta \sim 1.85$, making it broadly representative of moderately dense regions within the CMZ.

We perform SED fitting at this position under four different configurations: (a) Full-band coverage using 70, 160, 350, 450, and 850~$\mu$m (the configuration adopted throughout this work); (b) The same as (a), but excluding 850~$\mu$m; (c) The same as (a), but excluding 70~$\mu$m; (d) Excluding both 70 and 850~$\mu$m bands.
In each case, we employ MCMC sampling to estimate the posterior distributions of $T_\mathrm{dust}$ and $\beta$.
Figure \ref{fg_SED_corner} shows the resulting joint posterior distributions. 
When the full 70--850 $\mu$m range is used, the posteriors for both parameters are nearly Gaussian and exhibit a well-defined elliptical shape in the $T_\mathrm{dust}$–$\beta$ space, indicating minimal degeneracy. 
In contrast, excluding the measurement at either the shortest or longest wavelength introduces stronger degeneracies and significantly enlarged uncertainties in $T_{\rm dust}$ and $\beta$. 
This can also be seen in the comparison of the results of SED fittings under configurations (a) and (d) (Figure \ref{fg_SED_fitting}).
This highlights the importance of including both short- and long-wavelength data, which provide complementary constraints to break the $T_\mathrm{dust}$–$\beta$ degeneracy and improve the fit reliability.

To further evaluate the degree of $T_\mathrm{dust}$–$\beta$ degeneracy in our sample, we examine the fitted temperature and emissivity index distributions across all six clouds in our study (Figure \ref{fg_T_beta_distribution}).
These distributions do not exhibit strong anti-correlations, suggesting that $T_\mathrm{dust}$ and $\beta$ are reasonably decoupled.
This outcome supports the reliability of our fitting approach and indicates that our full-band data set is sufficient to mitigate the intrinsic degeneracy.


\subsection{Mid-IR Absorption Features}
\label{appendix:SED_abs}

\begin{figure}[htb!]
    \includegraphics[width=8.cm]{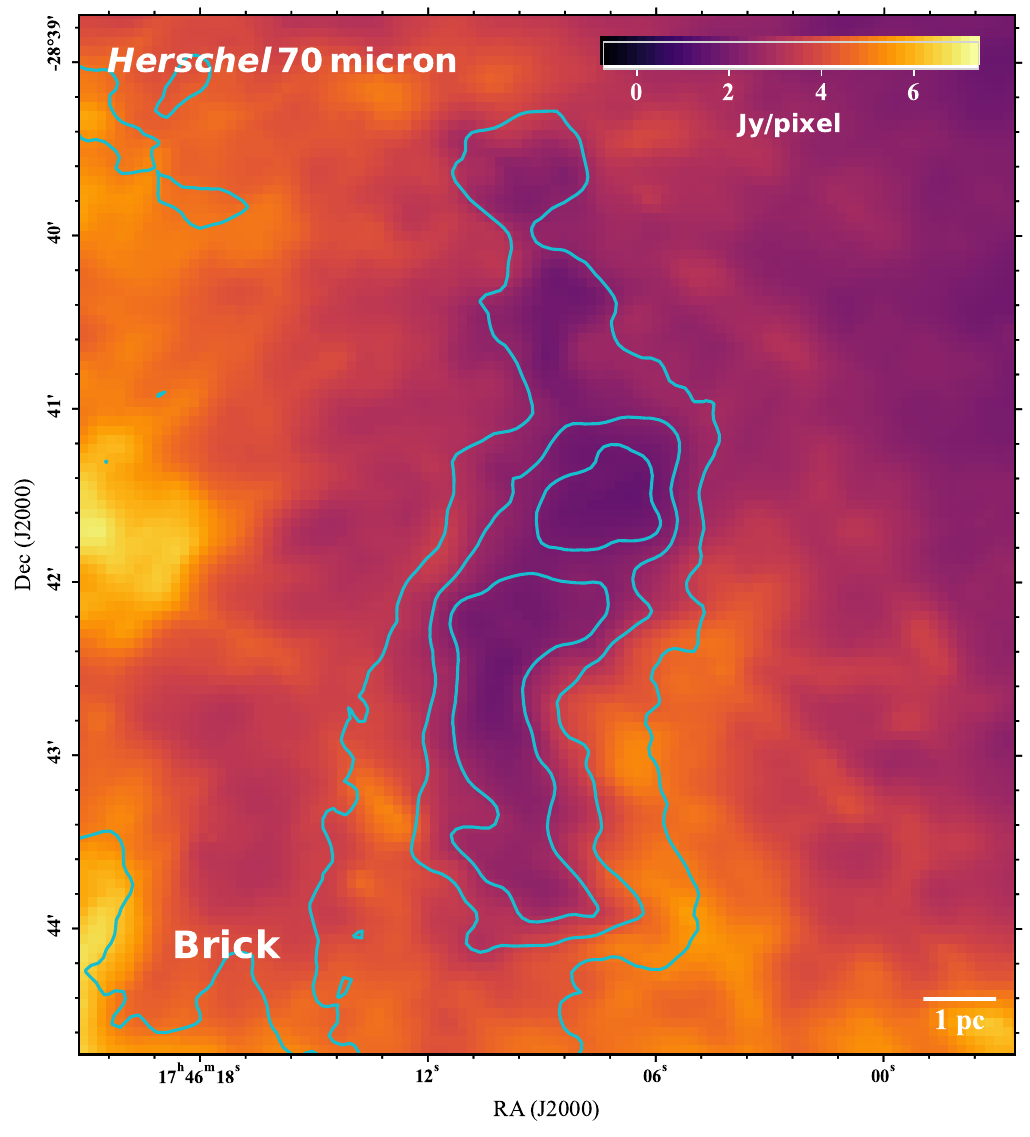} \\
    \caption{
    Herschel 70 $\mu$m intensity map of the Brick, showing the strong mid-IR absorption against the bright background emission.
    Cyan contours represent the JCMT/SCUBA-2 450 $\mu$m continuum at 5, 10, and 15 $\sigma$ levels, tracing the dense structure of the cloud.
    }
\label{fg_brick_70}
\end{figure}

When a cold and dense molecular cloud lies in front of a bright mid-IR background, the observed mid-IR emission can show absorption features. 
In the CMZ, abundant warm diffuse gas produces strong mid-IR background emission, so such absorption structures are seen toward several clouds, including the Brick (Figure \ref{fg_brick_70}), the 20 km s$^{-1}$ clouds, and the 50 km s$^{-1}$ clouds.
In these cases, the observed spectrum involves a more complex radiative transfer process: thermal emission from the warm diffuse background is attenuated by the foreground cold cloud, while the cloud itself also emits thermally.
This can be expressed as
\begin{equation}
    \frac{S_{\nu}}{\Omega} = {\rm B}_{\nu}(T_{\rm bg})(1-e^{-\tau_{\rm bg}})e^{-\tau_{\rm cloud}}+{\rm B}_{\nu}(T_{\rm cloud})(1-e^{-\tau_{\rm cloud}}),
\label{eq_SED_abs}
\end{equation}
where the subscripts `bg' and `cloud' denote the background warm diffuse gas and the foreground cold cloud, respectively.

\begin{figure}[htb!]
    \hspace{-0.3cm}
        \includegraphics[width=8.8cm]{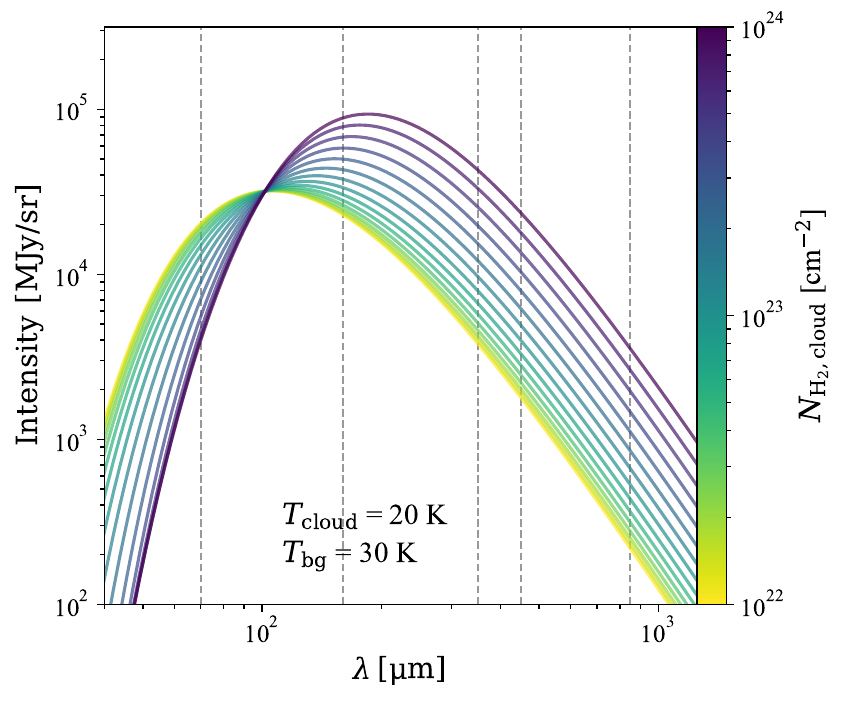}
    \caption{
    Example SEDs that include both a foreground cold cloud (20 K) and a background component of warm diffuse gas (30 K).
    The background component is assumed to have a column density of $3\times10^{22}\:{\rm cm^{-2}}$.
    The color scale indicates the column density of the foreground cloud increasing from $10^{22}$ to $10^{24}\,{\rm cm^{-2}}$.
    Vertical dashed lines mark the five bands used for the SED fitting in this work.
    }
\label{fg_SED_abs_model}
\end{figure}

Figure \ref{fg_SED_abs_model} illustrates an example SED that includes both a foreground cold cloud and a background warm diffuse gas. 
We assume the background diffuse gas has a temperature of $T_{\rm bg}\sim30{\rm K}$ and a column density of $N_{\rm H_2,bg}\sim3\times10^{22}\:{\rm cm^{-2}}$, while the foreground cold cloud has a temperature of $T_{\rm cloud}\sim20{\rm K}$. 
A dust emissivity index of $\beta=1.8$ is adopted for both components.
These typical values are consistent with our SED fitting results and with previous studies \citep[e.g.,][]{Battersby2025ApJ...984..156B,Tangyuping2021MNRAS.505.2392T,Molinari2011ApJ...735L..33M}.
As the column density of the foreground cloud increases, the overall spectrum gradually shifts from background-dominated to foreground-dominated.
During this transition, the intensity at 70 $\mu$m decreases markedly, while the emission between 160 and 850 $\mu$m increases, consistent with the observed trends in these bands.

\begin{figure*}[htb!]
    \hspace{-0.2cm}
    \includegraphics[width=18cm]{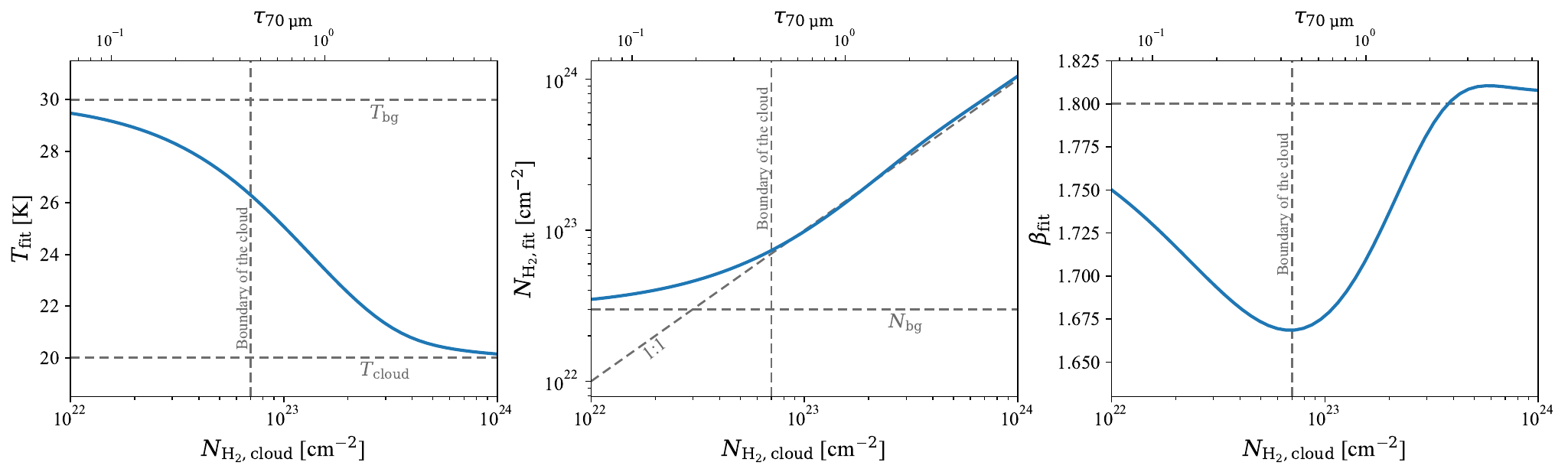} \\
    \caption{
    Results of fitting the two-component model SEDs (Equation \ref{eq_SED_abs} and Figure \ref{fg_SED_abs_model}) with a single modified blackbody (Equation \ref{eq1}).
    The panels show the fitted temperature $T_{\rm fit}$, column density $N_{\rm H_2, fit}$, and emissivity index $\beta_{\rm fit}$ as functions of the true column density of the foreground cloud, $N_{\rm H_2,cloud}$.
    Horizontal Gray dashed lines mark the adopted parameters for the background and foreground components.
    The vertical dashed line indicates the lowest column density adopted as the cloud boundary in our work, and the diagonal dashed line in the middle panel shows the 1:1 relation for comparison.
    }
\label{fg_SED_abs_test}
\end{figure*}

To evaluate how such two-component structures affect our single–temperature SED fitting, we performed tests by fitting the model SEDs in Figure \ref{fg_SED_abs_model} with a single modified blackbody function (Equation \ref{eq1}).
This quantifies the deviations between the fitted and true parameters in the presence of absorption and line-of-sight temperature contrast. Figure~\ref{fg_SED_abs_test} presents the results of these tests.
Overall, the radiative transfer approaches the form of a single modified blackbody when the foreground cloud is either very optically thin or very optically thick.
In the former case, the observed SED mainly traces the dust properties of the warm diffuse component, while in the latter, it reflects the dense cold cloud.
Although both components contribute to the dust property maps, their spatial mixture is complex.
The largest departures from a single-graybody description occur near the boundaries of dense clouds, where the foreground column is substantial but the mid-IR optical depth is only moderate.

Based on our tests, we highlight the following points:
\begin{itemize}
\setlength{\itemsep}{4pt}
\setlength{\parskip}{0pt}
\setlength{\parsep}{0pt}
    \item Mid-infrared absorption features appear only in a small fraction of the CMZ, primarily toward dense molecular clouds such as the Brick, the 20 km s$^{-1}$ clouds, the 50 km s$^{-1}$ clouds, and Sgr~B1-off.
    \item The fitted column density is not seriously affected in our work.
    The dense, cold component dominates the total column by more than an order of magnitude, and our multi-band SED fitting includes sufficient long-wavelength (optically thin) coverage. 
    Consequently, the derived N-PDFs, mass estimates, and subsequent scaling relations are robust to this effect.
    \item The fitted dust temperature represents a line-of-sight average.
    For regions where the foreground column density is significant but not optically thick in the mid-IR ($\tau_{70\,\mu{\rm m}} \sim 0.3\text{–}2$), the derived temperature tends to fall between that of the cold cloud and the warm background.
    In contrast, when no foreground cloud is present or when the foreground is extremely opaque at mid-IR ($N_{\mathrm{H}_2} \gtrsim 2\text{–}3\times10^{23}\,{\rm cm^{-2}}$), the temperature can be well approximated by a single graybody.
    Therefore, the derived temperature maps are not largely affected across most CMZ regions, except near the boundaries of a few dense cold clouds.
    \item The emissivity index beta is also impacted at a minor level.
    For regions where $\tau_{70\,\mu{\rm m}}$ lies between about 0.1 and 2, the fitted $\beta$ can be underestimated by up to $\sim0.1$.
    Interpreting beta variations may therefore require multi-component SED modeling to mitigate this bias.
    \item Our tests adopt a simplified configuration with a uniform, isothermal background.
    In reality, the CMZ background emission is not purely external but partly arises within the same complex, warm structures.
    These tests should be regarded as providing order-of-magnitude guidance on the impact of mid-IR absorption.
\end{itemize}

\section{Fitting Results of CH$_3$CN and NH$_3$ for Core Temperature Estimation}
\label{appendix:tem_fit}

We performed CH$_3$CN and its isotope CH$_3^{13}$CN line fitting for the hot core of Sgr B1-off (aka. cloud e), the 20 km/s cloud, and Sgr C. In the framework of XCLASS, the source size (i.e., beam filling factor), excitation temperature, column density, velocity offset, and line width are free parameters. We didn't constrain the isotope ratio but set it as a free parameter. When extracting spectra, we consider two cases of aperture: 0.4 arcsec (equivalent to the beam size) and 1 arcsec (similar to the hot core size). Both cases have a good fitting result by XCLASS, as shown in red and yellow line models, respectively. 
We consider the former case as the spectra at the continuum peak, corresponding to a higher temperature. But the 1 arcsec case is considered better as an averaged temperature for the hot core. Therefore, we finally adopted 1-arcsec temperature in our analyses. 

\begin{figure*}[htb!]
    \hspace{-1.0cm}
    \begin{tabular} { p{9cm} }
        \includegraphics[width=18cm]{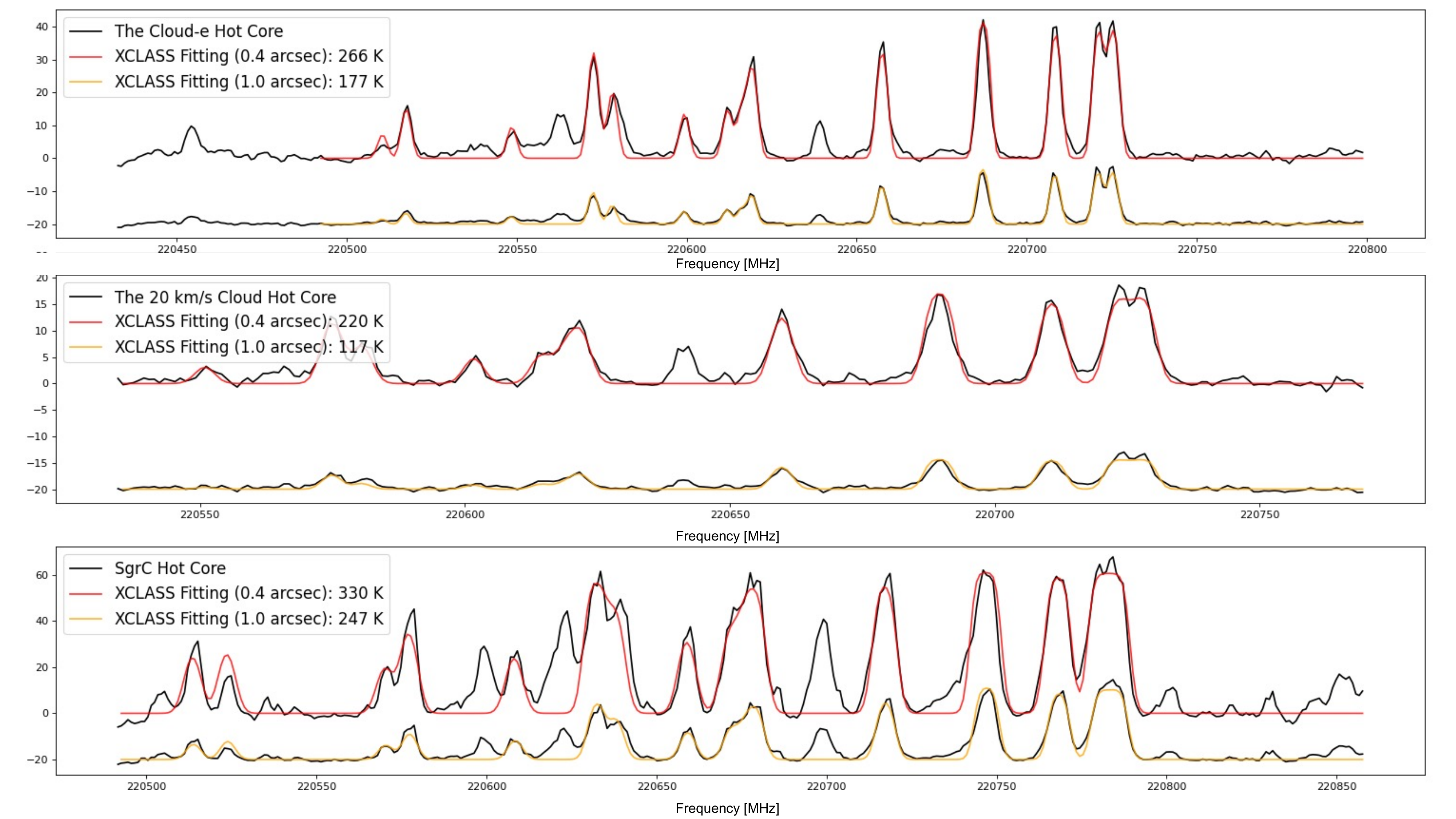} \\
        \hspace{0.45cm}\includegraphics[width=17cm]{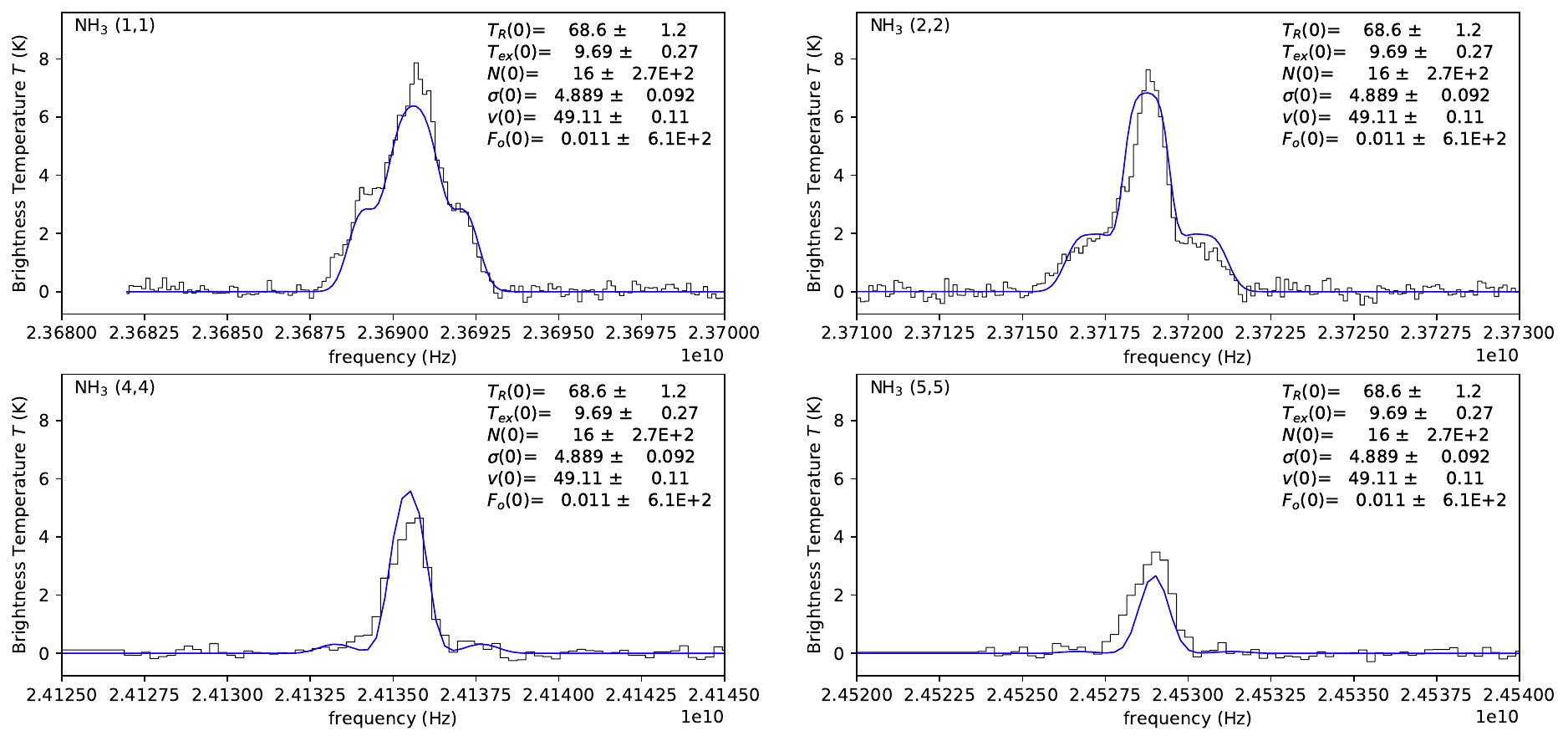} \\
    \end{tabular}
    \caption{
    The spectral line fitting results of CH$_3$CN and NH$_3$.
    The CH$_3$CN spectra (top panels) are fitted for the hot cores in the Sgr B1-off, the 20 km s$^{-1}$ cloud, and Sgr C, while the NH$_3$ spectra (bottom panels) are fitted for the 50 km s$^{-1}$ cloud.
    }
\label{fg_Trot}
\end{figure*}

In the 50 km/s cloud, the CH$_3$CN is not detected robustly and is not associated with the continuum peak. 
So we chose ammonia (NH$_3$), which is detected throughout the cloud region. Adopting NH$_3$ metastable lines (1,1), (2,2), (4,4), and (5,5), the rotation temperature ($T_R$) can be fitted using the ammonia line model in \texttt{pyspeckit} \citep{2022AJ....163..291G}.

\end{appendix}

\bibliography{sample701}{}
\bibliographystyle{aasjournalv7}



\end{document}